\documentclass[a4paper,fleqn,usenatbib,useAMS]{mn2e}
\usepackage[applemac]{inputenc}
\usepackage[draft=false, colorlinks=true, allcolors=blue]{hyperref}
\usepackage{graphicx}	% Including figure files
\usepackage{amsmath}	% Advanced maths commands
\usepackage{amssymb}	% Extra maths symbols
\usepackage{mathrsfs}

\newcommand{\percent}{\ensuremath{\mathrm{per\, cent}}}
\newcommand{\mh}{\ensuremath{h^{-1} \, \mathrm{M_\odot}}}
\newcommand{\mpch}{\ensuremath{h^{-1} \, \mathrm{Mpc}}}

\newcommand{\mean}[1]{\bar{#1}}
\newcommand{\amean}[1]{\langle#1\rangle}
\newcommand{\diff}{\mathrm{d}}
\newcommand{\vect}[1]{\ensuremath{\mathbf{#1}}}

\newcommand{\OmtrueIII}{\ensuremath{0.02_{-0.10}^{+0.11}}}
\newcommand{\StrueIII}{\ensuremath{0.001_{-0.007}^{+0.005}}}
\newcommand{\wtrueIII}{\ensuremath{-0.03_{-0.16}^{+0.16}}}
\newcommand{\OmWLIII}{\ensuremath{-0.06_{-0.10}^{+0.12}}}
\newcommand{\SWLIII}{\ensuremath{-0.023_{-0.008}^{+0.007}}}
\newcommand{\wWLIII}{\ensuremath{0.02_{-0.16}^{+0.19}}}
\newcommand{\OmWLcIII}{\ensuremath{-0.09_{-0.09}^{+0.11}}}
\newcommand{\SWLcIII}{\ensuremath{-0.023_{-0.008}^{+0.007}}}
\newcommand{\wWLcIII}{\ensuremath{0.09_{-0.18}^{+0.18}}}

\newcommand{\OmtrueIV}{\ensuremath{-0.0_{-0.008}^{+0.009}}}
\newcommand{\StrueIV}{\ensuremath{0.0_{-0.0004}^{+0.0003}}}
\newcommand{\wtrueIV}{\ensuremath{0.001_{-0.014}^{+0.011}}}
\newcommand{\OmtruescattertwentyIV}{\ensuremath{0.003_{-0.008}^{+0.010}}}
\newcommand{\StruescattertwentyIV}{\ensuremath{0.001_{-0.003}^{+0.003}}}
\newcommand{\wtruescattertwentyIV}{\ensuremath{0.002_{-0.013}^{+0.012}}}
\newcommand{\OmtruemargtwentyIV}{\ensuremath{-0.001_{-0.008}^{+0.009}}}
\newcommand{\StruemargtwentyIV}{\ensuremath{-0.0002_{-0.0004}^{+0.0004}}}
\newcommand{\wtruemargtwentyIV}{\ensuremath{0.002_{-0.013}^{+0.012}}}

\newcommand{\OmWLIV}{\ensuremath{-0.078_{-0.008}^{+0.009}}}
\newcommand{\SWLIV}{\ensuremath{-0.0291_{-0.0004}^{+0.0004}}}
\newcommand{\wWLIV}{\ensuremath{0.055_{-0.015}^{+0.014}}}
\newcommand{\OmWLscattertwentyIV}{\ensuremath{-0.055_{-0.009}^{+0.010}}}
\newcommand{\SWLscattertwentyIV}{\ensuremath{-0.047_{-0.004}^{+0.004}}}
\newcommand{\wWLscattertwentyIV}{\ensuremath{0.055_{-0.014}^{+0.013}}}
\newcommand{\OmWLmargtwentyIV}{\ensuremath{-0.077_{-0.008}^{+0.009}}}
\newcommand{\SWLmargtwentyIV}{\ensuremath{-0.0297_{-0.0004}^{+0.0004}}}
\newcommand{\wWLmargtwentyIV}{\ensuremath{0.047_{-0.014}^{+0.013}}}

\newcommand{\OmWLcIV}{\ensuremath{-0.113_{-0.008}^{+0.009}}}
\newcommand{\SWLcIV}{\ensuremath{-0.0329_{-0.0004}^{+0.0004}}}
\newcommand{\wWLcIV}{\ensuremath{0.091_{-0.015}^{+0.014}}}
\newcommand{\OmWLcscattertwentyIV}{\ensuremath{-0.076_{-0.009}^{+0.010}}}
\newcommand{\SWLcscattertwentyIV}{\ensuremath{-0.066_{-0.004}^{+0.004}}}
\newcommand{\wWLcscattertwentyIV}{\ensuremath{0.100_{-0.016}^{+0.014}}}
\newcommand{\OmWLcmargtwentyIV}{\ensuremath{-0.112_{-0.009}^{+0.009}}}
\newcommand{\SWLcmargtwentyIV}{\ensuremath{-0.0337_{-0.0004}^{+0.0004}}}
\newcommand{\wWLcmargtwentyIV}{\ensuremath{0.089_{-0.015}^{+0.014}}}

\usepackage[T1]{fontenc}
\usepackage{ae,aecompl}

\usepackage{newtxtext,newtxmath}
\title[Baryons bias cluster cosmology]{How baryons can significantly bias cluster
  count cosmology}

\author[S.N.B. Debackere et al.]{
  Stijn N.B. Debackere\thanks{Contact e-mail:
    \href{mailto:debackere@strw.leidenuniv.nl}{debackere@strw.leidenuniv.nl}
  },
  Joop Schaye,
  Henk Hoekstra
  \\
  Leiden Observatory, Leiden University, PO Box 9513, NL-2300 RA
  Leiden, The Netherlands
  \\}

\date{Last updated --; in original form --}

\pubyear{2021}

\begin{document}\label{firstpage}
\pagerange{\pageref{firstpage}--\pageref{lastpage}}
\maketitle

\begin{abstract}
  We quantify two main pathways through which baryonic physics biases
  cluster count cosmology. We create mock cluster samples that
  reproduce the baryon content inferred from X-ray observations. We
  link clusters to their counterparts in a dark matter-only universe,
  whose abundances can be predicted robustly, by assuming the dark
  matter density profile is not significantly affected by baryons. We
  derive weak lensing halo masses and infer the best-fitting
  cosmological parameters $\Omega_\mathrm{m}$,
  $S_8=\sigma_8(\Omega_\mathrm{m}/0.3)^{0.2}$, and $w_0$ from the mock
  cluster sample. We find that because of the need to accommodate the
  change in the density profile due to the ejection of baryons, weak
  lensing mass calibrations are only unbiased if the concentration is
  left free when fitting the reduced shear with NFW profiles. However,
  even unbiased total mass estimates give rise to biased cosmological
  parameters if the measured mass functions are compared with
  predictions from dark matter-only simulations. This bias dominates
  for haloes with $m_\mathrm{500c} < 10^{14.5} \, \mh$. For a stage
  IV-like cluster survey without mass estimation uncertainties, an
  area $\approx 15000 \, \mathrm{deg^2}$ and a constant mass cut of
  $m_\mathrm{200m,min} = 10^{14} \, \mh$, the biases are
  $-11 \pm 1 \, \percent$ in $\Omega_\mathrm{m}$,
  $-3.29 \pm 0.04 \, \percent$ in $S_8$, and $9 \pm 1.5 \, \percent$
  in $w_0$. The statistical significance of the baryonic bias depends
  on how accurately the actual uncertainty on individual cluster mass
  estimates is known. We suggest that rather than the total halo mass,
  the (re-scaled) dark matter mass inferred from the combination of
  weak lensing and observations of the hot gas, should be used for
  cluster count cosmology.
\end{abstract}


%%%%%%%%%%%%%%%%%%%%%%%%%%%%%%%%%%%%%%%%%%%%%%%%%%

%%%%%%%%%%%%%%%%% BODY OF PAPER %%%%%%%%%%%%%%%%%%

\section{Introduction}\label{sec:introduction}
Clusters of galaxies are sensitive probes of structure formation in a universe
where structure forms hierarchically, because they are still actively forming.
Their abundance in a given volume as a function of mass and redshift contains a
wealth of information about the formation history of the Universe, i.e. its
total amount of matter, how clustered it is, and how its accelerated expansion
changed in time \citep[e.g.][]{Allen2011}. The fact that the cluster abundance
drops exponentially with increasing mass enables precise constraints on the
underlying cosmology, but it also necessitates accurate mass calibrations
\citep[e.g.][]{Evrard1989, Bahcall1997}.

Linking observed cluster number counts to the theoretical expectation for a
given cosmology requires a well-defined cluster selection function and an
accurately calibrated mass--observable relation. These requirements are not
independent, as \citet{Mantz2019} illustrated how the selection function also
plays an important role in constraining the assumed scaling relations between
the observable mass proxy and the true mass near the survey mass limit. All
current abundance studies account for these effects in their analysis
\citep{Mantz2010, DeHaan2016, Bocquet2019, DESCollaboration2020}. While the
cluster selection function is a crucial part of the cosmological analysis, it
also depends on the cluster detection method and is thus survey-specific. Here,
we will assume that the completeness of the sample can be modelled perfectly and
focus solely on the calibration of the mass--observable relation.

To convert the observed cluster mass proxy, e.g. the Sunyaev-Zel'dovich (SZ)
detection significance, into a mass, we need the mass--observable scaling
relation. The mass--observable relation cannot be predicted robustly from first
principles, since it relies on complex galaxy formation physics. Calibrating
this scaling relation requires unbiased mass estimates for a subset of the
cluster sample. Consequently, it is generally calibrated using weak lensing
observations as they probe the total matter content of the cluster
\citep[e.g.][]{VonderLinden2014, VonderLinden2014a, Hoekstra2015,
  Schrabback2018, Dietrich2019a, McClintock2019}.

\citet{Kohlinger2015} have shown the dramatic reduction in the statistical
uncertainties and systematic errors in cluster mass estimates from an idealized
weak lensing analysis due to the expected increase in area and background galaxy
number density of stage IV-like surveys such as
\emph{Euclid}\footnote{\url{https://www.euclid-ec.org}} and the Rubin
Observatory Legacy Survey of Space and Time (\emph{LSST})
\footnote{\url{https://www.lsst.org/}}. However, the accuracy of weak lensing
mass calibrations remains an open question, especially in the presence of
baryons. \citet{Bahe2012} investigated the mass bias inferred from weak lensing
observations in dark matter-only (DMO, i.e. gravity-only) simulations, finding
cluster masses to be biased low by $\approx 5 \, \percent$ due to deviations of
the cluster density profile from the assumed Navarro-Frenk-White (NFW, see
\citealt{Navarro1996}) shape in the cluster outskirts. Similarly,
\citet{Henson2017} found a bias of up to $\approx 10 \, \percent$ in
hydrodynamical simulations. The main conclusion from these studies is that we
need to correct weak lensing-derived masses for the lack of spherical symmetry
of the observed halo using virtual observations of simulated haloes \citep[see
e.g.][]{Dietrich2019a}. \citet{Lee2018a} used hydrodynamical simulations to show
that while these effects are certainly important, the coherent suppression of
the inner halo density profile due to baryonic physics also matters. The impact
of this effect on cluster number count cosmology has not been isolated so far.

Simulations indicate that baryons significantly change the density
profiles of haloes when comparing them to their matched DMO
counterparts \citep[e.g. ][]{Velliscig2014, VanDaalen2014a, Lee2018a}.
In hydrodynamical simulations, baryonic effects lower the halo mass,
$m_\mathrm{200c}$, at the $\lesssim 5 \, (1) \, \percent$ level for
cluster-sized haloes with
$m_\mathrm{200c} > 10^{14} \, (10^{14.5}) \, \mh$ compared to the same
halo mass in a gravity-only simulation\footnote{We define the
  spherical overdensity masses as the mass contained inside the
  physical radii $r_\mathrm{\Delta c}(z)$, $r_\mathrm{\Delta m}(z)$
  that enclose an average density of
  $\amean{\rho} = \Delta \rho_\mathrm{crit}(z)$,
  $\amean{\rho} = \Delta \rho_\mathrm{m}(z) =
  \Omega_\mathrm{m}\rho_\mathrm{crit}(z=0)(1+z)^3$, respectively,
  where $\rho_\mathrm{crit}(z=0) = 3H_0^2/(8\pi G)$. That is,
  $m_\mathrm{\Delta c} = 4 / 3 \pi \Delta \rho_\mathrm{crit}(z)
  r_\mathrm{\Delta c}^3(z)$ and
  $m_\mathrm{\Delta m}(z) = 4/3 \pi \Delta \Omega_\mathrm{m}
  \rho_\mathrm{crit}(z=0)(1+z)^3 r_\mathrm{\Delta m}^3(z)$.},
$m_\mathrm{200c,dmo}$. \citep[e.g.][]{Sawala2013, Cui2014,
  Velliscig2014, Martizzi2014a, Bocquet2016, Castro2020}. Hence, we
should not expect cluster density profiles to follow the NFW shape,
especially since baryons are preferentially ejected outside
$r \approx r_\mathrm{500c}$, where weak lensing observations reach
their optimal signal-to-noise ratio. \citet{Balaguera-Antolinez2013}
have investigated the impact of the halo mass change due to baryons on
cluster count cosmology, but they did not include the effect of weak
lensing mass calibrations. To isolate the effect of the change in the
halo density profile due to baryons, we generated idealized, spherical
clusters that consist of dark matter and hot gas that reproduces the
observed cluster X-ray emission, thus bypassing the large inherent
uncertainties associated with the assumed subgrid models in
hydrodynamical simulations. These models allow us to study the bias in
the inferred halo masses for a standard, mock weak lensing analysis
that assumes NFW density profiles.

With the cluster masses determined, the number counts as a function of mass and
redshift need to be linked to the underlying cosmology. Generally, the
cosmology-dependence of the halo mass function is taken from N-body (i.e.
gravity-only) simulations due to the need to simulate large volumes to obtain
complete samples of clusters at high masses for a range of cosmologies and
because of the large uncertainties associated with baryonic physics. Hence, the
aforementioned change in halo density profile also complicates the link between
observed haloes and their DMO equivalents whose abundance we can predict
robustly \citep[e.g.][]{Cui2014, Cusworth2014, Velliscig2014}. Since stage
IV-like surveys will reliably detect clusters down to halo masses of
$m_\mathrm{500c} \approx 10^{14} \, \mh$, this disconnect between observed and
DMO haloes will need to be taken into account in their cosmological analyses.

In this paper, we investigate the impact of baryonic effects on cluster number
count cosmology. We build a self-consistent, phenomenological model that links
idealized clusters whose baryon content matches that inferred from X-ray
observations, to their DMO equivalents (Section~\ref{sec:hm}). Our linking
method relies on the assumption that the cluster dark matter profile does not
change significantly due to the presence of baryons. Then, we determine the
cluster masses from mock weak lensing observations assuming NFW profiles with
either fixed or free concentration--mass relations
(Section~\ref{sec:observations}). We show how the resulting mass biases impact
cosmological parameters for different surveys in Section~\ref{sec:cosmo}. In
Section~\ref{sec:aperture_masses} we explore the performance of aperture masses,
which do not depend as sensitively on the halo density profile. The change in
the inner density profile due to baryonic effects affects aperture masses less
strongly than deprojected masses, resulting in a closer, but still not perfect,
correspondence to the equivalent DMO halo masses. We compare our findings to the
literature in Section~\ref{sec:discussion} and conclude in
Section~\ref{sec:conclusions}.

\section{Halo mass model}\label{sec:hm}

We construct an idealized model for the halo matter content as a function of
halo mass that incorporates observations for the baryonic component. We modify
the model used in our previous work, where we used a halo model to study the
impact of baryonic physics on the matter power spectrum \citep{Debackere2020}.
The goal here is to obtain halo density profiles that reproduce the observed hot
gas density profiles from galaxy clusters while at the same time constraining
their abundance through the mass of their equivalent DMO halo and the halo mass
function calibrated with DMO simulations. This will allow us to
self-consistently study the impact of baryonic physics on cluster number count
cosmology.

\subsection{Linking observed and DMO haloes}\label{sec:hm:dmo_linking}

In short, a halo contains dark matter and baryons. In this paper, we assume that
the latter consists entirely of hot gas, and we ignore the stars since they
contribute only a small fraction ($\approx 1 \, \percent$) of the total mass and
since the satellite component, which dominates the stellar mass, approximately
follows an NFW density profile, similarly to the dark matter \citep[see
e.g.][]{VanderBurg2015a}. The main assumption required to link observed haloes
to their equivalent DMO haloes is that the presence of baryons does not
significantly affect the bulk of the dark matter. If this is the case, the dark
matter of the observed halo will follow the density profile of the equivalent
DMO halo, but with a lower normalization, i.e.
\begin{equation}
\label{eq:m_dmo_500c}
m_\mathrm{dmo}({<}r) =  \frac{m_\mathrm{dm}({<}r)}{1 - \Omega_\mathrm{b} /
  \Omega_\mathrm{m}} \, .
\end{equation}
We can convert the observationally inferred total halo mass $m({<}r)$ to the DM
mass at the same radius using the observed baryon fraction $f_\mathrm{bar}(r)$,
\begin{equation}
  \label{eq:m_dm_500c}
  m_\mathrm{dm}({<}r) = (1 - f_\mathrm{bar}(r)) m({<}r) \, .
\end{equation}
Imposing an NFW profile so that
\begin{equation}
  \label{eq:m_dmo_nfw}
  m_\mathrm{dmo}({<}r; c(m_\mathrm{200m,dmo}, z)) = 4 \pi \int\limits_0^r \rho_\mathrm{NFW}(r; c(m_\mathrm{200m,dmo}, z)) r^2 \diff r \, ,
\end{equation}
and combining Eqs.~\eqref{eq:m_dmo_500c} and~\eqref{eq:m_dm_500c}, yields
\begin{equation}
\label{eq:m500c_obs_to_dmo}
m_\mathrm{dmo}({<}r; c(m_\mathrm{200m,dmo}, z)) = \frac{1 - f_\mathrm{bar}(r)}{1 - \Omega_\mathrm{b} / \Omega_\mathrm{m}}m({<}r) \, .
\end{equation}
These relations fully determine the dark matter density profile and the
equivalent DMO halo corresponding to the observed halo relying solely on the
observed baryon fraction, the inferred total halo mass, and an assumed density
profile for the DMO halo. We adopt an NFW density profile \citep{Navarro1996} for
the equivalent DMO halo and the median concentration--mass relation,
$c(m_\mathrm{200m,dmo}, z)$, for relaxed haloes without scatter of
\citet{Correa2015c}. \citet{Brown2020} have shown that this relation accurately
predicts the concentration of simulated DMO haloes in observationally allowed
$\Lambda$CDM cosmologies. Explicitly, we assume that the dark matter of the
observed halo has the same scale radius as the equivalent DMO halo, but a density
that is a factor of $1 - \Omega_\mathrm{b} / \Omega_\mathrm{m}$ lower.

Eq.~\eqref{eq:m_dmo_500c} will not hold in detail since the dark matter does
react to the presence of baryons \citep[e.g.][]{Gnedin2004, Duffy2010,
  Schaller2015}. However, in the \textsc{OWLS} \citep{Schaye2010} and
cosmo-\textsc{OWLS} simulations \citep{LeBrun2014a} the dark matter mass enclosed
within $r_\mathrm{200c}$ increases by $< 1 \, \percent$ due to contraction for
all halo masses that we include in our analysis \citep{Velliscig2014}. Hence, by
not accounting for the contraction of the dark matter, we may overestimate the
true equivalent DMO halo mass by up to $\approx 1 \, \percent$, since
$m_\mathrm{dm}({<}r)/m_\mathrm{dmo}({<}r) > 1 - \Omega_\mathrm{b} /
\Omega_\mathrm{m}$. However, this effect will be smaller than the bias due to
missing baryons for the abundant low-mass clusters
($m_\mathrm{500c} \lesssim 10^{14.5} \, \mh$) that are missing a significant
fraction of the cosmic baryons.

\subsection{Including observations of baryons}\label{sec:hm:gas_observations}

\begin{figure}
  \centering \includegraphics[width=\columnwidth]{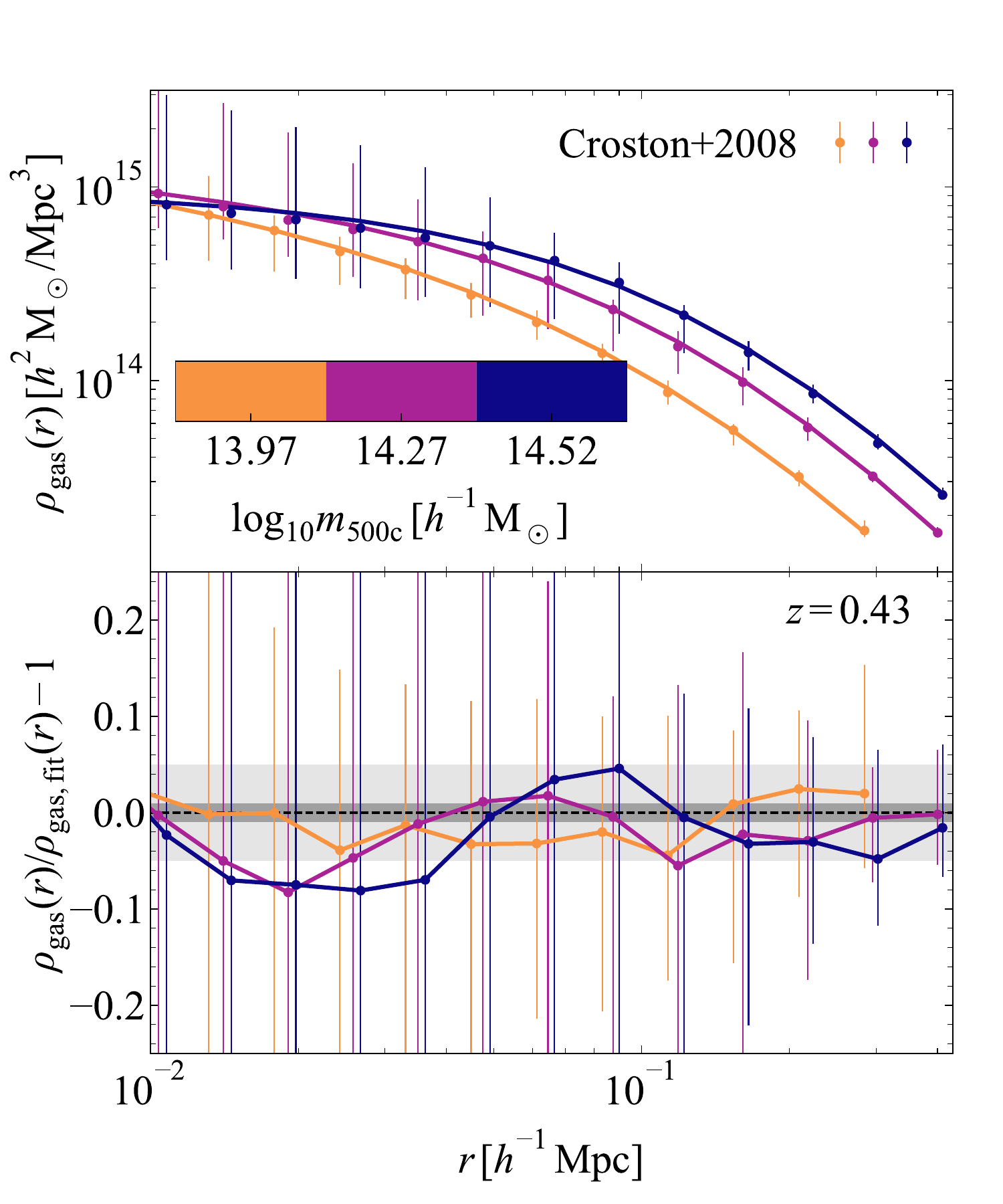}
  \caption{\emph{Top panel:} The median hot gas density profiles, evolved
    self-similarly to $z=0.43$, with their 16th and 84th percentile scatter for
    the halo mass-binned density profiles from \citet{Croston2008} (coloured
    circles). We also show the model gas density profiles inferred from fitting
    the halo baryon fractions (coloured lines). \emph{Bottom panel:} The ratio
    between the observed hot gas density profiles and our best-fitting model. We
    recover the observed profile at the $\approx 5 \, \percent$ level for most
    of the radial range, which is well within the scatter of the
    object-to-object scatter for individual mass bins.}\label{fig:rho_r}
\end{figure}
To determine the baryonic component of our model, we only require a fit to the
hot gas density profiles inferred from the observed X-ray surface brightness of
galaxy clusters. For a detailed description of how the X-ray surface brightness
is converted into the density profile, we refer to Section 3 of
\citet{Debackere2020}. In short, the X-ray surface brightness is fit with a
spherically symmetric, collisionally ionized electron plasma of temperature $T$
and metallicity $Z$. Assuming mass abundances for hydrogen, helium and metals,
we then convert the electron number density into a mass density profile. The
halo masses for each cluster can then be determined from the hot gas density and
temperature profiles under the assumption of hydrostatic equilibrium. We use
observations from the Representative XMM-Newton Cluster Structure Survey
(REXCESS, \citealt{Bohringer2007}) because the clusters constitute a local,
high-quality, and volume-limited sample, representative of the local X-ray
cluster population. Since the survey is not flux-limited, the sample suffers
less from the well-known cool-core bias for X-ray cluster samples
\citep{Chon2017a}. However, the dynamical state of REXCESS clusters still
differs from that of SZ selected samples \citep[which suffer less from biases
due to their approximate mass selection, see e.g.][]{Rossetti2016}. We evolve
the inferred density profiles self-similarly to extrapolate to higher redshifts.
In self-similar evolution, density profiles evolve with redshift as
$\rho(z) \propto E^2(z)=\Omega_\mathrm{m}(1+z)^3 + \Omega_\Lambda$
\citep{Kaiser1986}. Consequently, masses defined with respect to the critical
density of the Universe remain constant. In the top panel of
Fig.~\ref{fig:rho_r}, we show the median of the $m_\mathrm{500c}$-binned
observed hot gas density profiles, $\rho_\mathrm{gas}(r|m, z=0.43)$, evolved
self-similarly to $z=0.43$ (the mean redshift of both the SPT and DES
calibration samples, see \citealt{Dietrich2019a} and
\citealt{DESCollaboration2020}, respectively), and the 16th and 84th percentile
range from the REXCESS data of \citet{Croston2008}.

Our procedure for obtaining the gas density profiles and corresponding cluster
masses, relies on a couple of assumptions that we now justify. First, in linking
the gas density profiles inferred from X-ray observations to the cluster masses,
we have assumed hydrostatic equilibrium. This assumption implies that our
resulting masses are lower limits on the true cluster masses since observations
and simulations suggest that halo masses inferred from X-ray observations and
hydrostatic equilibrium are underestimated by $\approx 15-30 \, \percent$
\citep[e.g. ][]{Mahdavi2013, VonderLinden2014a, Hoekstra2015, Medezinski2018,
  Barnes2020, Herbonnet2020}. Looking at Eq.~\eqref{eq:m500c_obs_to_dmo}, the
mass ratio $m({<}r) / m_\mathrm{dmo}({<}r)$, whose bias we want to study,
depends inversely on the inferred dark matter fraction at $r$,
$1 - f_\mathrm{bar}(r)$. If the observed cluster were not in hydrostatic
equilibrium, the fixed overdensity radius would increase along with the halo
mass. If the halo baryon fraction increases with radius outside
$r_\mathrm{500c}$ (which is a valid assumption, see e.g.
\citealt{Vikhlinin2006b}), the resulting enclosed baryon fraction would be
higher than the one derived assuming hydrostatic equilibrium. In this case, the
true mass ratio between the observed halo and its corresponding DMO halo,
$m({<}r) / m_\mathrm{dmo}({<}r)$, would be lower than our value inferred
assuming hydrostatic equilibrium. Hence, our model provides an upper bound to
the minimum possible mass ratio bias in Eq.~\eqref{eq:m500c_obs_to_dmo} due to
missing baryons.

Second, we have assumed that the hot gas density profiles evolve self-similarly
with redshift. There is observational evidence that the redshift scaling of the
cluster hot outer gas density profile is indeed close to self-similar
\citep[e.g.][]{McDonald2017}.

\subsection{Fitting the gas density profiles}\label{sec:hm:fits}

In \citet{Debackere2020}, we constructed halo density profiles by fitting beta
profiles to the galaxy cluster gas density profiles inferred from the observed
X-ray emission. While this is certainly a valid approach, we take a different
route here. In our previous work, we had to enforce steeper slopes for the
observationally unconstrained outer hot gas density profile so that haloes did
not exceed the cosmic baryon fraction. However, while this fine-tuning process
ensures that the halo baryon fraction reaches the cosmic value at a fixed
radius, it then gradually declines further out. Since we wish to ensure that the
halo baryon fraction converges to the cosmic value in the halo outskirts, we
decided not to fit the gas density profile, but the halo baryon fraction
instead:
\begin{equation}
  \label{eq:fbar_r}
  f_\mathrm{bar}(r| m, z) = \frac{m_\mathrm{bar}({<}r | m, z)}{m_\mathrm{bar}({<}r |m, z) + m_\mathrm{dm}({<}r |m, z)} \, ,
\end{equation}
where $m_\mathrm{bar}({<}r|m,z)$ and $m_\mathrm{dm}({<}r|m,z)$ are the enclosed
baryonic and dark matter mass within $r$ for a halo of mass $m$ at redshift $z$,
respectively. We can enforce the convergence to the cosmic baryon fraction in
the halo outskirts by choosing a functional form for $f_\mathrm{bar}(r)$ that
asymptotes to $\Omega_\mathrm{b} / \Omega_\mathrm{m}$.

\begin{figure}
  \centering \includegraphics[width=\columnwidth]{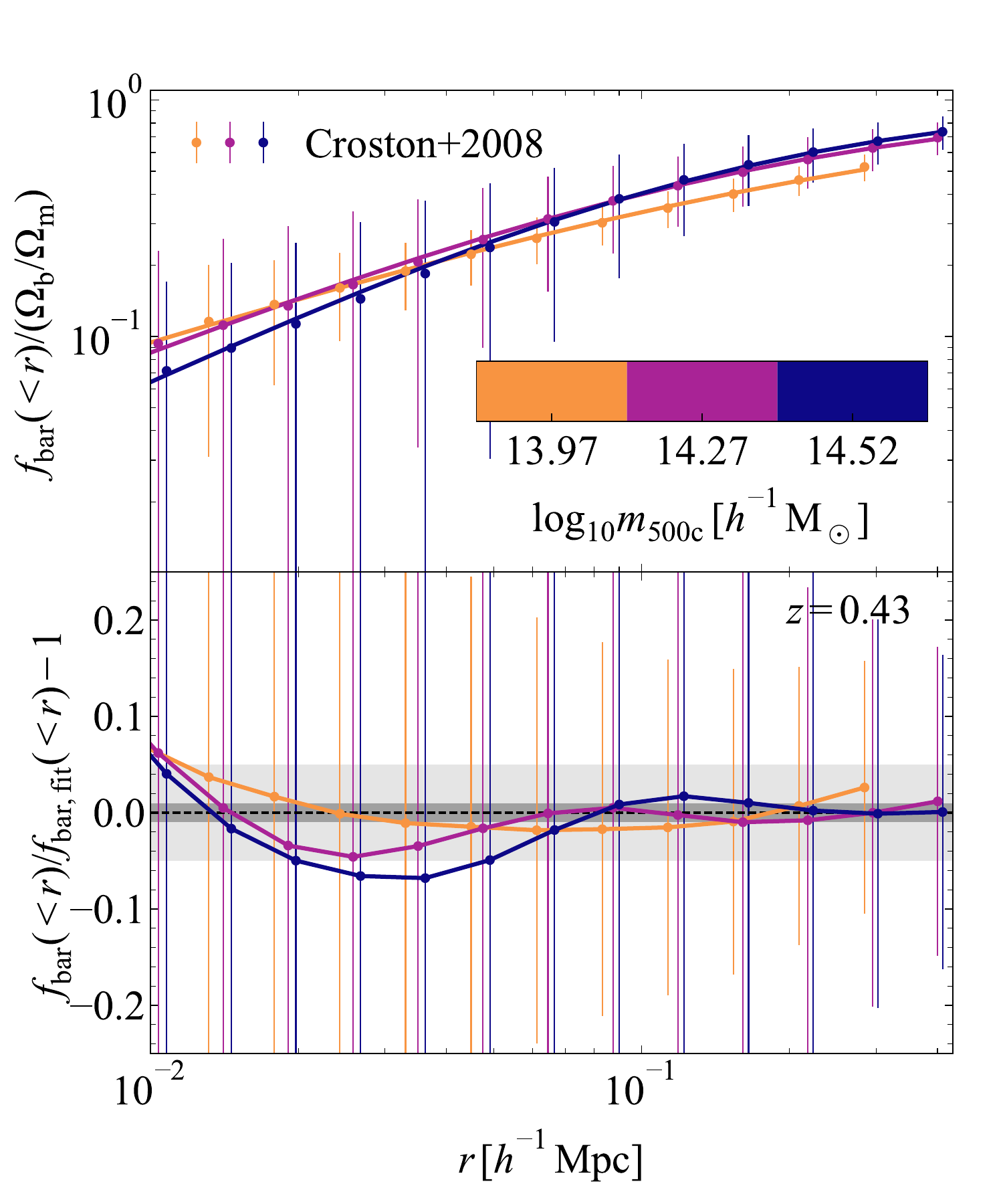}
  \caption{\emph{Top panel:} The enclosed baryon fraction as a function of
    radius for the median, mass-binned hot gas density profiles, evolved
    self-similarly to $z=0.43$, including their 16th and 84th percentile scatter
    from \citet{Croston2008} (coloured circles). The dark matter mass is
    obtained by subtracting the gas mass from the inferred total halo mass. We
    assume the dark matter follows an NFW profile with a scale radius determined
    by the equivalent DMO halo that accounts for all cosmic baryons within
    $r_\mathrm{500c}$, i.e.
    $m_\mathrm{dmo}({<}r_\mathrm{500c}) = 1 / (1 - \Omega_\mathrm{b} /
    \Omega_\mathrm{m}) m_\mathrm{dm}({<}r_\mathrm{500c})$. Our best-fitting
    model assuming Eq.~\eqref{eq:fbar_fun} is shown as the coloured lines.
    \emph{Bottom panel:} The ratio between the inferred enclosed baryon fraction
    from X-ray observations and our best-fitting model. We recover the correct
    baryon fractions at the $\approx 5 \, \percent$ level for all radii and halo
    masses, which is well within the scatter of the object-to-object scatter for
    individual mass bins.}\label{fig:fbar_r}
\end{figure}

We construct the enclosed baryon fraction profiles from the observed gas density
profiles $\rho_\mathrm{gas}(r|m, z)$ from the REXCESS data of
\citet{Croston2008}. For each cluster, we determine the dark matter mass at
$r_\mathrm{500c}$ using Eq.~\eqref{eq:m_dm_500c} and the NFW scale radius by
solving Eq.~\eqref{eq:m500c_obs_to_dmo}, assuming the hot gas accounts for all
the halo baryons. Then, we obtain $f_\mathrm{bar}(r|m,z)$ from
Eq.~\eqref{eq:fbar_r}. We show the halo baryon fraction inferred from the
observations, also evolved self-similarly to $z=0.43$, in the top panel of
Fig.~\ref{fig:fbar_r}.

The baryonic density profiles can be recovered by taking the derivative of the
enclosed baryonic mass profile (we drop the $m$ and $z$ dependence)
\begin{align}
  \label{eq:rho_bar}
  \nonumber
  \rho_\mathrm{bar}(r) & = \frac{1}{4\pi r^2}\frac{\diff m_\mathrm{bar}({<}r)}{\diff r} \\
  \nonumber
                       & = \frac{1}{4\pi r^2}\frac{\diff}{\diff r} \left( \frac{f_\mathrm{bar}(r)}{1 - f_\mathrm{bar}(r)} m_\mathrm{dm}({<}r) \right) \\
                       & = \frac{f_\mathrm{bar}^\prime(r) m_\mathrm{dm}({<}r)}{4\pi r^2 (1 - f_\mathrm{bar}(r))^2} + \frac{f_\mathrm{bar}(r)}{1 - f_\mathrm{bar}(r)} \rho_\mathrm{dm}(r) \, ,
\end{align}
where $^\prime \equiv \mathrm{d}/\mathrm{d}r$. For outer boundary conditions
$\lim_{r \to \infty} f_\mathrm{bar}(r) = \Omega_\mathrm{b} / \Omega_\mathrm{m}$
and $\lim_{r \to \infty} f_\mathrm{bar}^\prime (r) = 0$, it is clear that the
baryonic density profile will follow the dark matter in the halo outskirts. In
fact, the total matter profile, $\rho_\mathrm{bar} + \rho_\mathrm{dm}$, will
approach the equivalent DMO halo profile, since
$\rho_\mathrm{dm}(r) = (1 - \Omega_\mathrm{b} / \Omega_\mathrm{m})
\rho_\mathrm{dmo}(r)$. This is exactly what is found in simulations when
comparing the halo-matter cross-correlation (which traces the average halo
density profile for a given mass) between DMO and hydrodynamical simulations
\citep{VanDaalen2014a}.

In this paper we will assume the baryon fraction goes to zero at small radii for
simplicity. Different functional behaviours, for instance including a central
increase in the baryon fraction that captures the stellar contribution, are also
possible. However, we are interested in studying the change in the cluster weak
lensing signal due to the inclusion of baryons. Since the lensing analysis
usually excludes the central regions, and the central galaxy would only
contribute $\lesssim 1 \, \percent$ of the total halo mass \citep[see
e.g.][]{Zu2015}, we can safely neglect its contribution. We assume the profile
\begin{equation}
  \label{eq:fbar_fun}
  f_\mathrm{bar}(r|m, z) = \frac{\Omega_\mathrm{b} / \Omega_\mathrm{m}}{2}\left( 1 + \tanh\left( \frac{\log_{10}r - \log_{10}r_\mathrm{t}(m,z)}{\alpha(m,z)} \right) \right) \, ,
\end{equation}
which gives
\begin{equation}
  \label{eq:dfbardr_fun}
  f_\mathrm{bar}^\prime(r|m, z) = \frac{\Omega_\mathrm{b} / \Omega_\mathrm{m}}{2 \ln(10) \alpha(m, z) r} \cosh^{-2}\left( \frac{\log_{10}(r / r_\mathrm{t}(m,z))}{\alpha(m,z)} \right) \, ,
\end{equation}
where $r_\mathrm{t}(m,z)$ determines where the increase in the baryon fraction
turns over and $\alpha(m, z)$ sets the sharpness of the turnover ($\alpha \gg 1$
is smooth, $\alpha \ll 1$ is sharp). We show the best-fitting
$f_\mathrm{bar}(r|m,z)$ profiles to the REXCESS data, assuming
Eqs.~\eqref{eq:rho_bar},~\eqref{eq:fbar_fun}, and~\eqref{eq:dfbardr_fun}, in the
top panel of Fig.~\ref{fig:fbar_r}. In the bottom panel of
Fig.~\ref{fig:fbar_r}, we show the ratio of our model to the observations. We
are able to capture the observed behaviour at the $\approx 5 \, \percent$ level
for all halo masses and over most of the radial range. This accuracy is well
within the observed scatter of the individual gas density profiles. The benefit
of fitting the halo baryon fraction instead of the gas density, is that the
outer baryonic density automatically traces the dark matter, while accounting
for all of the cosmic baryons.

To extrapolate our model beyond the observed cluster masses and redshifts, we
scale the density profiles self-similarly and fit $r_\mathrm{t}(m, z)$ and
$\alpha(m, z)$, opting for the following $(m, z)$ dependencies
\begin{align}
  \label{eq:model_rt}
  \log_\mathrm{10}(r_\mathrm{t}/r_\mathrm{x})(m_\mathrm{x}, z) & = \tilde{r}(z) (\log_\mathrm{10}m_\mathrm{x} - \tilde{m}(z)) \\
  \label{eq:model_alpha}
  \alpha(m_\mathrm{x},z) & = \tilde{\alpha}(z) (\log_\mathrm{10}m_\mathrm{x} - \tilde{\mu}(z)) \, ,
\end{align}
where $[\tilde{r}(z), \tilde{m}(z), \tilde{\alpha}(z), \tilde{\mu}(z)]$ are free
fitting parameters at 10 redshift bins $z \in [0.1, 2]$ (we interpolate for
intermediate values of $z$) and $m_\mathrm{x}$ is the chosen halo mass
definition, $m_\mathrm{500c}$ in our case. The chosen linear behaviour captures
the average mass dependence of the fit parameters quite well, as we show in
Appendix~\ref{app:model_fits}.

We stress that the assumed functional form for $f_\mathrm{bar}(r|m,z)$
implicitly fixes the gas density profile in the halo outskirts. To account for
different outer gas density profiles, we also fit the halo baryon fractions
inferred from the 16th and 84th percentiles of the hot gas density profiles in
Fig.~\ref{fig:rho_r}. To ensure that these fits bracket the median profile
results for all masses and redshifts, we fix $\alpha(m, z)$ to the best-fitting
behaviour of the median profiles, and leave $r_\mathrm{t}(m, z)$ free to vary.
These different profile behaviours can quantify the effect of higher and lower
outer gas densities, which are difficult to constrain observationally, on the
inferred halo masses from weak lensing observations.

\begin{figure}
  \centering
  \includegraphics[width=\columnwidth]{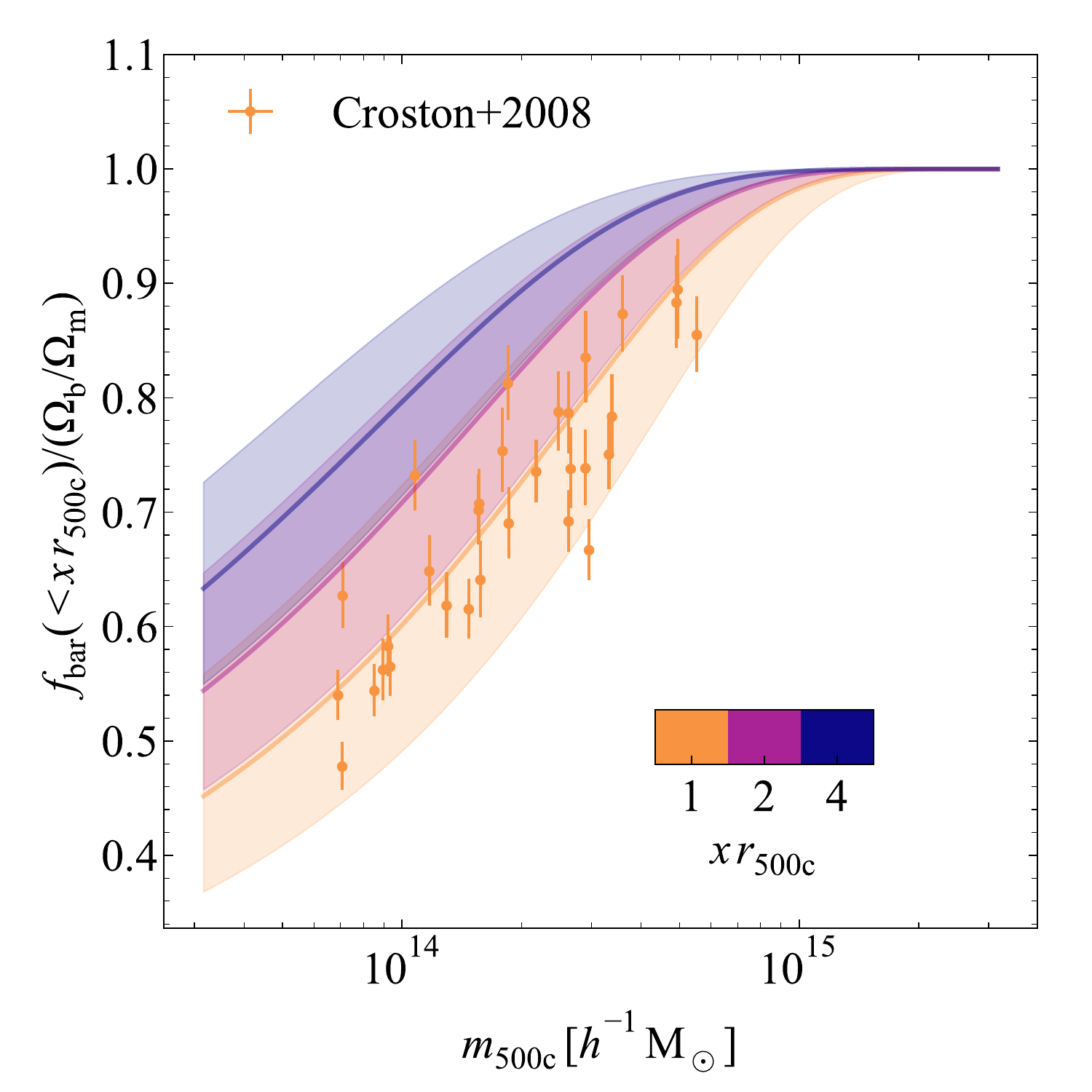}
  \caption{The REXCESS X-ray hydrostatic gas fractions as a function of halo mass
    from \citet{Croston2008}. The median
    \(f_{\mathrm{bar}}(<xr_\mathrm{500c})-m_{\mathrm{500c}}\) relations (thick,
    coloured lines) and the 16th to 84th percentile ranges (shaded regions) from
    our model fits to the inferred radial gas fractions of the observed density
    profiles are shown. We also show the extrapolated enclosed gas fractions at
    larger radii than observed.}\label{fig:obs_gas_fractions}
\end{figure}

We show the halo baryon fractions as a function of mass and for different outer
radii, in Fig.~\ref{fig:obs_gas_fractions}. We also show the gas fractions at
$r_\mathrm{500c}$ inferred from the REXCESS data. Our model closely reproduces
the median behaviour. The fits to the 16th and 84th percentiles of the hot gas
density profiles capture the full range of the observational uncertainty. Hence,
our model is fully representative of the REXCESS galaxy cluster population.

As a consequence of our chosen functional form for the radial profile of the halo
baryon fraction, Eq.~\eqref{eq:fbar_fun}, haloes with masses
$m_\mathrm{500c} \gtrsim 10^{15} \, \mh$ contain the cosmic baryon fraction
within $r_\mathrm{500c}$. In simulations, however, halo baryon fractions might
exceed the cosmic value at $r_\mathrm{500c}$ for these massive haloes since their
strong potential wells prevent the ejected baryons from leaving the halo (see
e.g. fig. A1 of \citealt{Velliscig2014} or fig. 2 of \citealt{Lee2018a}). The
REXCESS clusters are not massive enough to observe this behaviour. Moreover, even
if this were the case, the lower-mass haloes most tightly constrain the shape and
normalization of the halo mass function since they are more abundant.

Another possibly important effect is that at radii larger than $r_\mathrm{500c}$
the hot gas pressure might prevent further infall of cosmic baryons, lowering
the asymptotic baryon fraction below the cosmic value. Our mock weak lensing
observations are performed at scales $\approx r_\mathrm{500c}$ for the most
massive haloes, and should not be significantly affected by the gas distribution
in the halo outskirts. Our profiles assume that the baryon fraction asymptotes
to the universal fraction. If the baryon fraction at large radii were smaller
than assumed, then the true halo mass, $m_\mathrm{200m,true}$, would be lower
than our model prediction. In that case, the ratio
$m_\mathrm{200m,true} / m_\mathrm{200m,dmo}$ would be smaller than what we find,
since the linked DMO halo mass would remain the same. Hence, our model provides
an upper limit to the true mass ratio and, consequently, a lower limit on the
bias in the measured cosmological parameters from cluster counts. We stress that
Eq.~\eqref{eq:fbar_r} would be able to capture these behaviours if an
appropriate functional form is chosen.

In conclusion, our model accurately captures the baryonic content of the average
cluster population, since we fit it to the median halo mass-binned gas density
profiles inferred from cluster X-ray surface brightness profiles. This also
justifies our assumption of spherical symmetry, since deviations due to the
presence of substructure or triaxiality of individual haloes average out in a
stacked analysis if the cluster selection is unbiased. In
Section~\ref{sec:observations}, we will use our model to compare the halo masses
inferred from a mock weak lensing analysis to the true halo masses.

\section{Mock observational analysis}\label{sec:observations}

Mass calibrations of observed samples of clusters are carried out for a subset
of the sample for which weak lensing observations are available or follow-up
observations are made \citep[e.g. ][]{Applegate2014, Hoekstra2015,
  Schrabback2018, Dietrich2019a}. Different groups use different assumptions to
derive weak lensing masses. To minimize the statistical noise in the mass
determination of individual clusters due to the degeneracy between mass and
concentration \citep[see e.g.][]{Hoekstra2011}, one generally assumes a fixed
concentration (as in \citealt{Applegate2014} and \citealt{VonderLinden2014}) or
a concentration--mass relation from simulations (as in \citealt{Hoekstra2015,
  Schrabback2018, Dietrich2019a}). The weak lensing derived halo masses are then
used to calibrate a scaling relation between a survey observable mass proxy and
the weak lensing-derived halo mass.

Using our idealized halo model described in Section~\ref{sec:hm}, we can
generate mock weak lensing observables for clusters with realistic baryonic
density profiles. We investigate how accurately the aforementioned weak lensing
derived halo mass recovers the true halo mass in the presence of baryons and how
the best-fitting mass from the mock weak lensing observations compares to the
mass of the same halo in a gravity-only universe, for which we can reliably
predict the abundance. The mismatch between these masses determines the bias in
the cosmological parameters inferred from a cluster count cosmological analysis
as we will perform in Section~\ref{sec:cosmo}.

The observable of interest for weak lensing is the reduced shear
\begin{equation}
  \label{eq:g_T}
  g_\mathrm{T}(\theta) = \frac{\gamma_\mathrm{T}(\theta)}{1 - \kappa(\theta)} \, ,
\end{equation}
where $\kappa(\theta)=\Sigma(\theta)/\Sigma_\mathrm{crit}$ is the convergence,
$\gamma_\mathrm{T}(\theta)$ is the tangential shear, and $\Sigma_\mathrm{crit}$
is the critical surface mass density, defined as
\begin{align}
  \label{eq:sigma_crit}
  \Sigma_\mathrm{crit} & = \frac{c^2}{4 \pi G} \frac{1}{\beta D_\mathrm{l}}\, ,
\end{align}
where $D_\mathrm{l}$ and $\beta = \mathrm{max}(0, D_\mathrm{ls} / D_\mathrm{s})$
are the angular diameter distance between the observer and the lens, and the
lensing efficiency for a source at a distance $D_\mathrm{s}$ from the observer
and a distance $D_\mathrm{ls}$ behind the lens (which is negative for sources in
front of the lens), respectively.

For clusters, generally $\kappa \approx \gamma_\mathrm{T} \approx 0.01-0.1$ at
the scales probed with weak lensing observations
($0.5 \lesssim R \lesssim 5 \, \mpch$). Assuming a cosmological model, the
angular position, $\theta$, can be converted into a projected physical distance,
$R$, using the observed angular diameter distance, $D$, as $\theta = R / D$. The
tangential shear is given by
\begin{equation}
  \label{eq:gamma_T}
  \gamma_\mathrm{T}(R|m, z) = \frac{\mean{\Sigma}({<}R|m, z) - \Sigma(R|m, z)}{\Sigma_\mathrm{crit}} \, ,
\end{equation}
where $\Sigma(R|m, z)$ is the projected surface mass density profile for a halo
with mass $m$ at redshift $z$,
\begin{align}
  \label{eq:sigma}
  \nonumber
  \Sigma(R|m, z) & = \int_{-\infty}^{\infty} \rho(R, l|m, z) \diff l \\
                 & = 2 \int_R^\infty \diff r \, \rho(r|m ,z) \frac{r}{\sqrt{r^2 - R^2}} \, ,
\end{align}
which we compute with a Gauss-Jacobi quadrature to ensure convergence in the
presence of the singularity at $r=R$, and $\mean{\Sigma}({<}R|m, z)$ is the mean
enclosed surface mass density inside $R$
\begin{equation}
  \label{eq:sigma_mean}
  \mean{\Sigma}({<}R|m, z) = \frac{2}{R^2} \int_0^R \diff R^\prime \, R^\prime\Sigma(R^\prime|m, z) \, .
\end{equation}

The halo model described in Section~\ref{sec:hm} enters in Eqs.~\eqref{eq:sigma}
and~\eqref{eq:sigma_mean} through the total density profile
\begin{equation}
  \label{eq:rho_model}
  \rho(r|m, z) = \rho_\mathrm{dm}(r|m, z) + \rho_\mathrm{bar}(r|m, z) \, .
\end{equation}
Here, we obtain the normalization of the dark matter NFW density profile,
$\rho_\mathrm{dm}$, by taking the halo mass at $r_\mathrm{500c}$ and correcting
it for the gas fraction inferred from observations of the X-ray surface
brightness profiles of the REXCESS clusters (Eq.~\ref{eq:m_dm_500c}). We assume
that the dark matter has the same NFW scale radius as the equivalent DMO halo,
which can be derived by combining Eqs.~\eqref{eq:m_dmo_500c}
and~\eqref{eq:m_dmo_nfw}. The baryonic density profile, $\rho_\mathrm{bar}$, is
obtained by fitting Eq.~\eqref{eq:fbar_fun} to the radial baryon fraction
profiles inferred from observations.

\begin{figure}
  \centering \includegraphics[width=\columnwidth]{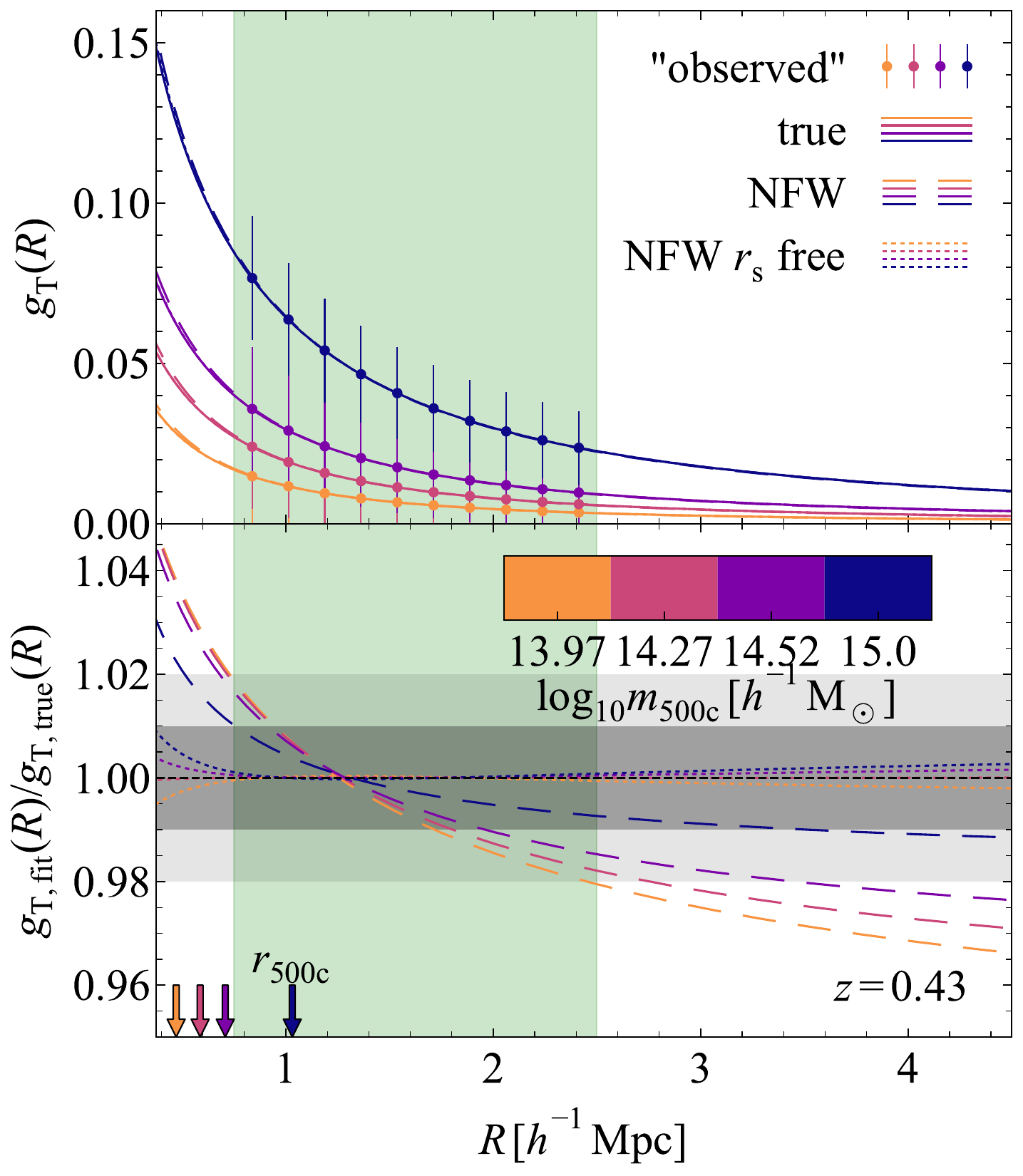}
  \caption{\emph{Top panel:} The reduced shear profiles for different halo mass
    bins (different colours) at $z=0.43$. The mock observations with
    uncertainties for a single halo set by an intrinsic galaxy shape noise of
    $\sigma_\mathrm{gal}=0.25$ and mean background galaxy density of
    $\mean{n}=10 \, \mathrm{arcmin^{-2}}$ are shown on top of the underlying
    true density profile (coloured dots and solid lines). The green shaded
    region indicates the fitting range for the mock weak lensing observations.
    The coloured arrows in the bottom panel indicate $r_\mathrm{500c}$ for the
    different halo mass bins. The best-fitting NFW profiles with fixed (free)
    scale radius, $r_\mathrm{s}$, are also shown as dashed lines (dotted lines).
    \emph{Bottom panel:} The ratio of the best-fitting NFW profiles to the true
    profiles. Leaving the NFW scale radius free results in accurate fits to the
    true profiles. Fixing the scale radius to a concentration--mass relation for
    DMO haloes overestimates the signal in the core, where baryons are missing,
    and underestimates the signal in the outskirts. The mismatch decreases with
    increasing halo mass as more massive haloes have higher baryon fractions
    within the fitting range.}\label{fig:g_T_fit}
\end{figure}

We show the reduced shear profiles for different halo mass bins in the top panel
of Fig.~\ref{fig:g_T_fit}. We have assumed a mean lensing efficiency
$\amean{\beta} = 0.5$ in Eq.~\eqref{eq:sigma_crit} (in agreement with the SPT
calibration sample; \citealt{Dietrich2019a}) to generate observations in 10
radial bins between $0.75 \, \mpch$ and $2.5 \, \mpch$ at $z=0.43$, (similar to
the mean redshift of the calibration samples for SPT and DES, $\amean{z} = 0.42$
and $\amean{z}=0.45$, respectively; \citealt{Dietrich2019a,
  DESCollaboration2020}). The observational uncertainty in the reduced shear due
to the intrinsic galaxy shape noise for each bin $R_i$ with bin size
$\Delta R_i$ decreases with the total number of galaxies in the bin, and is
taken to be
\begin{equation}
  \label{eq:shape_noise}
  \sigma_\mathrm{obs}^2 = \frac{\sigma_\mathrm{gal}^2}{2\pi\mean{n} R_i \Delta R_i} \, ,
\end{equation}
with the intrinsic galaxy shape noise $\sigma_\mathrm{gal} = 0.25$
\citep[e.g.][]{Hoekstra2000}, and the mean background galaxy number density
$\mean{n} = 10 \, \mathrm{arcmin^{-2}}$ \citep[similar to ][]{Dietrich2019a}. In
a stacked analysis the shape noise would decrease by a factor of $\sqrt{N}$,
where $N$ is the number of clusters in the stack. However, this would not affect
our best-fitting models since we do not include scatter in the mock
observational data. Our mock observations are overly optimistic in this sense.
However, given enough clusters, the derived mass--observable relation should
converge to the one we find. We choose radial bins within the range
$0.75 < R_i / (\mpch) < 2.5$ corresponding to angular sizes
$3.2 < \theta / \mathrm{arcmin} < 10.7$ at $z=0.43$ for a
\cite{PlanckCollaboration2018g} cosmology \citep[similar to ][]{Dietrich2019a}.
The inner radius corresponds to
$\approx 1.6 \, (0.5) \, r_\mathrm{500c}(z=0.43)$ for haloes of masses
$m_\mathrm{500c} = 10^{14} \, (10^{15.5}) \, \mh$. At smaller scales, cluster
miscentring and contamination become important. At larger scales, the
large-scale structure contributions to the surface mass density become
important. For different redshifts, we scale the radial range of the
observations by $(1+z)^{-1}$, i.e. $R_i(z) = 1.43 / (1 + z) R_i(z=0.43)$, to
ensure that we are not greatly exceeding $r_\mathrm{500c}(z)$ in the fitting
range.

The dashed lines in Fig.~\ref{fig:g_T_fit} indicate the best-fitting NFW profile
to the observed data points, assuming the median \citet{Correa2015c}
concentration--mass relation. We also show the resulting NFW profile when
leaving the scale radius, $r_\mathrm{s}$, free as the dotted lines.
Observationally, they would be difficult to distinguish from the true profile
because the difference due to baryons is negligible compared to the shape noise
of an individual cluster. The lower panel of Fig.~\ref{fig:g_T_fit} shows the
ratio between the best-fitting NFW reduced shear profiles and the true profiles.
Clearly, with currently attainable source background densities, we cannot
discern the true reduced shear profile from the best-fitting NFW profiles, which
would require \percent\ level precision for the shear measurements. We have
checked that even a stage IV-like survey with
$\mean{n} = 30 \, \mathrm{arcmin^{-2}}$ could only observe the difference
between the true density profile and the NFW fit with fixed concentration--mass
relation at the $\approx 2 \, \sigma$ level in a stack of $\mathcal{O}(10^4)$
clusters with $m_\mathrm{500c} > 10^{14} \mh$.

\begin{figure}
  \centering
  \includegraphics[width=\columnwidth]{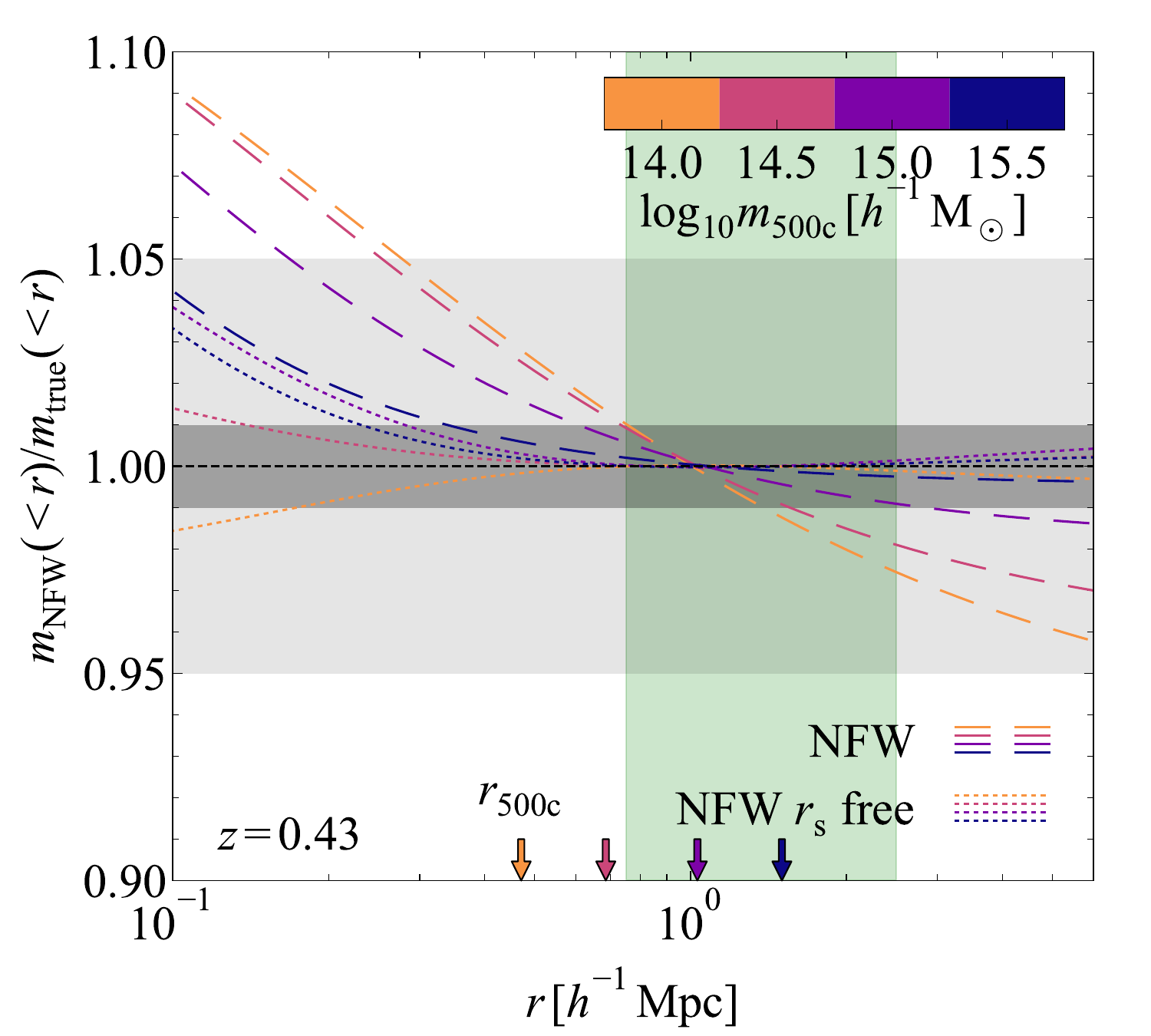}
  \caption{The ratio of the 3D enclosed total mass recovered from the
    best-fitting NFW profiles to the reduced shear with fixed and free scale
    radius, $r_\mathrm{s}$, to the true mass profile (dashed and dotted lines,
    respectively) for haloes of different masses at $z=0.43$. The green, shaded
    region indicates the radial range for the fit. The overdensity radii
    $r_\mathrm{500c}$ corresponding to the true density profiles are indicated
    with arrows. Fixing the concentration--mass relation of the NFW profile
    consistently overestimates (underestimates) the inner (outer) halo mass,
    where the baryonic mass is lower (higher) than the NFW prediction. Leaving
    the concentration of the NFW profile free removes the underestimation of the
    outer halo mass.}\label{fig:mr_NFW_vs_true}
\end{figure}

We obtain deprojected enclosed total halo mass profiles $m_\mathrm{NFW}({<}r)$
from the best-fitting NFW density profiles to the reduced shear. We show the
ratio between the NFW reconstructed enclosed halo mass with fixed and free scale
radius, $r_\mathrm{s}$, and the true halo mass in Fig.~\ref{fig:mr_NFW_vs_true}
for haloes with masses
$m_\mathrm{500c} = 10^{14}, 10^{14.5}, 10^{15}, 10^{15.5} \, \mh$. The results
of both fitting methods are generally within $\approx 5 \, \percent$ of the true
enclosed mass profiles for all halo masses we show. However, fixing the
concentration--mass relation of the NFW density results in substantially more
biased halo mass estimates. The best-fitting NFW profile is determined by the
fitting range of the observations and minimizes the $\chi^2$ error by balancing
the over- and underestimation of the true profile, as can be seen in the bottom
panel of Fig.~\ref{fig:g_T_fit}. Since feedback processes redistribute the
baryons to larger scales, the best-fitting NFW profiles consistently
overestimate the halo mass internal to the minimum radius of the fit. Moreover,
since the NFW profile cannot capture the more rapidly increasing baryonic mass
towards the halo outskirts, the outer halo mass is consistently underestimated.
This behaviour is general: the inner radius of the observational fitting range
approximately determines the physical scale at which the inferred total
deprojected halo masses are unbiased. For radii progressively smaller (larger)
than the inner fitting radius, total deprojected masses are overestimated
(underestimated) with increasing amplitude.

\begin{figure}
  \centering
  \includegraphics[width=\columnwidth]{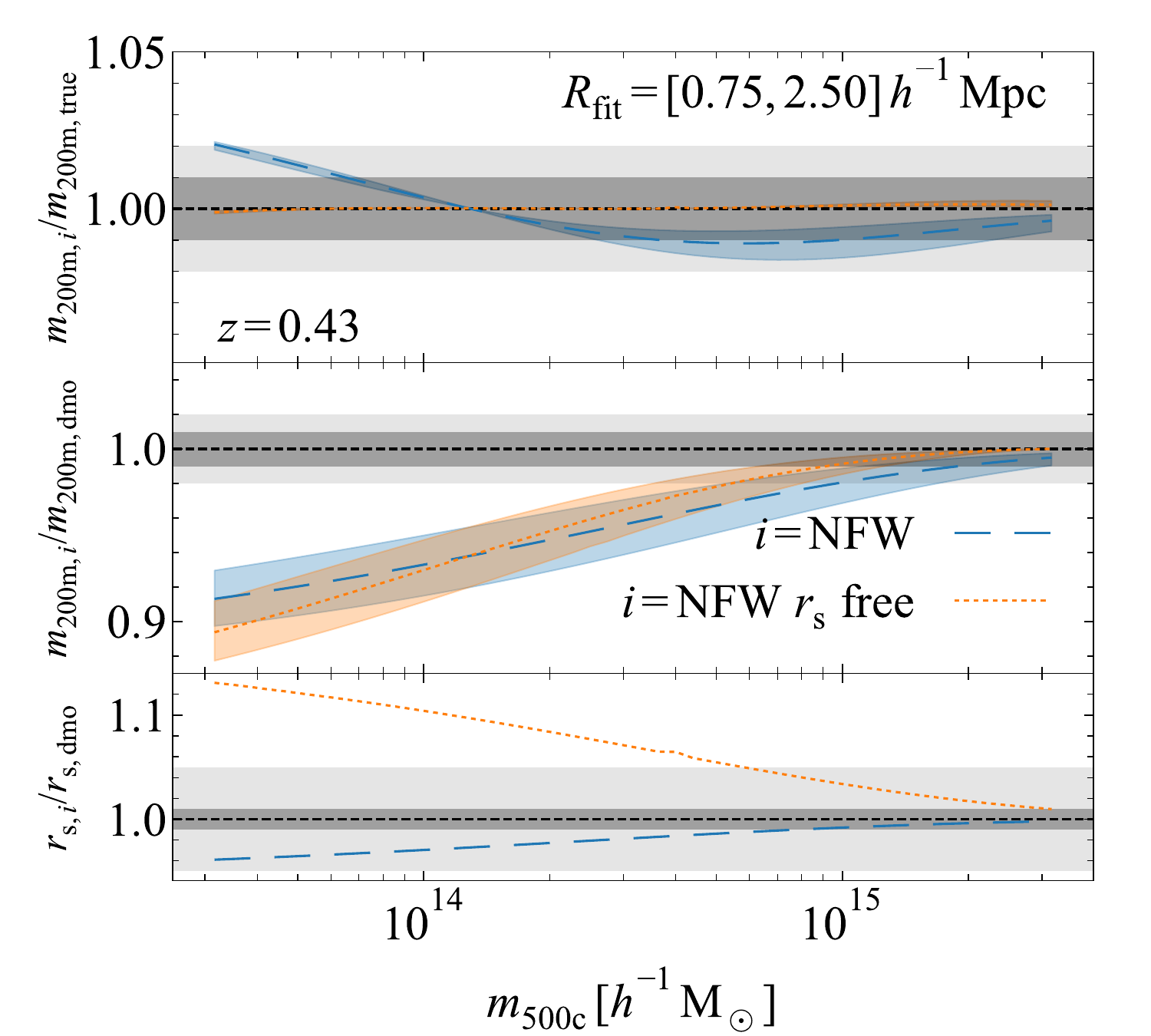}
  \caption{\emph{Top panel:} The ratio of the 3D enclosed total overdensity
    mass, $m_\mathrm{200m,NFW}$, inferred from the best-fitting NFW profiles to
    the reduced shear, to the true halo mass $m_\mathrm{200m,true}$ as a
    function of $m_\mathrm{500c}$. The dashed and dotted lines show the mass
    ratio $m_\mathrm{200m,NFW}/m_\mathrm{200m,true}$ for the best-fitting NFW
    density profiles with fixed and free scale radius for a fitting
    range of $R_\mathrm{fit}=[0.75-2.5] \, \mpch$, respectively. Fixing the
    scale radius results in biased estimates for $m_\mathrm{200m,true}$, leaving
    the scale radius free removes this bias. \emph{Middle panel:} The ratio of
    the inferred halo mass $m_\mathrm{200m,NFW}$ to the equivalent dark
    matter-only halo mass $m_\mathrm{200m,dmo}$ as a function of
    $m_\mathrm{500c}$. The resulting mass ratios are biased for both mass
    determination methods, since the missing halo baryons bias
    $m_\mathrm{200m,true}$ with respect to $m_\mathrm{200m,dmo}$. \emph{Bottom
      panel:} The ratio of the scale radius of the best-fitting NFW profile to
    the true dark matter-only NFW scale radius $r_\mathrm{s,dmo}$. Leaving the
    scale radius free results in larger values, since the baryons are less
    centrally concentrated than the dark matter.}\label{fig:m200m_ratio_NFW}
\end{figure}

This bias can be reduced, however, by leaving the NFW scale radius as a free
parameter. The inner halo mass will still be biased, but the extra freedom
allows for practically unbiased outer halo mass estimates (see
Fig.~\ref{fig:mr_NFW_vs_true}). This behaviour is clearly visible in the top
panel of Fig.~\ref{fig:m200m_ratio_NFW}, where we show the ratio
$m_\mathrm{200m,NFW}/m_\mathrm{200m,true}$ for both fitting methods. The bottom
panel of Fig.~\ref{fig:m200m_ratio_NFW} shows how $r_\mathrm{s}$ needs to
increase with respect to the true value to capture the less centrally
concentrated halo baryons. However, this is not possible when fixing the
concentration--mass relation, resulting in overestimated (underestimated) masses
when $r_\mathrm{200m,true} \lesssim (\gtrsim) 1 \, \mpch$ (at
$\approx 1 \, \mpch$ and $z=0.43$, the enclosed mass estimates are unbiased for
our chosen fitting range, this corresponds to
$m_\mathrm{500c} \approx 10^{14.1} \, \mh$).

The halo mass $m_\mathrm{200m,NFW}$ from the best-fitting NFW density profile
can be used to obtain unbiased estimates of the true halo mass
$m_\mathrm{200m,true}$ if the concentration--mass relation is left free.
However, for cluster abundance studies, the mass of interest is not
$m_\mathrm{200m,true}$ of the observed halo, but the halo mass of the equivalent
DMO halo, $m_\mathrm{200m,dmo}$. All calibrated fitting functions and emulators
of the halo mass function are obtained from DMO simulations \citep[e.g.
][]{Tinker2008, McClintock2018, Nishimichi2019, Bocquet2020}, since the matter
distribution in hydrodynamical simulations depends sensitively on the assumed
``subgrid'' physics recipes required to model the complex galaxy formation
processes \citep[e.g.][]{Velliscig2014}.

We show the ratio $m_\mathrm{200m,NFW} / m_\mathrm{200m,dmo}$ as a function of
halo mass $m_\mathrm{500c}$ in the middle panel of
Fig.~\ref{fig:m200m_ratio_NFW}. We do not show the ratio
$m_\mathrm{200m,true}/m_\mathrm{200m,dmo}$ for the actual halo mass since it
matches the relation for the best-fitting NFW density profile with a free scale
radius (shown as the dotted line) almost exactly (the halo mass
$m_\mathrm{200m,NFW}$ is nearly unbiased when the NFW scale radius is left
free). The suppression of the true halo mass with respect to the equivalent DMO
halo stems from the missing halo baryons within $r_\mathrm{200m,true}$. Fixing
the concentration--mass relation of the NFW density profile (shown as the dashed
line) results in biases similar to leaving the NFW scale radius free, except for
the small modulation due to the mass bias in $m_\mathrm{200m,NFW}$ with respect
to $m_\mathrm{200m,true}$ (see the top panel of Fig.~\ref{fig:m200m_ratio_NFW}).
As mentioned earlier, this bias stems from the chosen radial fitting range for
the weak lensing observations. Decreasing (increasing) the inner fitting radius
shifts the crossover between over- and underestimated $m_\mathrm{200m,true}$ to
lower (higher) halo masses and changes the overall amplitude of the bias.
Remarkably, for low-mass haloes ($m_\mathrm{500c} \lesssim 10^{14.1} \, \mh$),
the overestimation of $m_\mathrm{200m,true}$ when fixing the concentration--mass
relation results in a less biased estimate of $m_\mathrm{200m,dmo}$. However, we
would preferably not rely on more biased estimates of the true halo mass to
obtain less biased cosmological parameters.

We find a slightly stronger suppression in the ratio
$m_\mathrm{200m,true} / m_\mathrm{200m,dmo}$ in our model compared to
cosmo-\textsc{OWLS} (for $r_\mathrm{s}$ free we find $> 1 \, \percent$
suppression for $m_\mathrm{500c} \lesssim 10^{15} \, \mh$ compared to
$m_\mathrm{500c} \lesssim 10^{14.5} \, \mh$ in \citealt{Velliscig2014}). The
reason for this is twofold. First, we do not include a stellar component in our
model. Since stars are more centrally concentrated than the hot gas, the NFW
fits in cosmo-\textsc{OWLS} perform slightly better in the inner regions,
capturing an extra $\approx 1 \, \percent$ of the total halo mass there and
reducing the mass ratio bias. Second, in cosmo-\textsc{OWLS} contraction of the
dark matter component due to the baryons at these halo masses slightly reduces
the bias since more dark matter mass is included in the central regions than we
are accounting for in Eq.~\eqref{eq:m_dmo_500c}. However, for
$m_\mathrm{500c} \gtrsim 10^{14} \, \mh$, the dark matter contraction increases
the enclosed halo mass ratio $m_\mathrm{dm}({<}r) / m_\mathrm{dmo}({<}r)$ in
Eq.~\eqref{eq:m_dmo_500c} by only $\lesssim 1 \, \percent$ \citep[see fig. 3
of][]{Velliscig2014}. For $m_\mathrm{500c} \lesssim 10^{14} \, \mh$, the dark
matter actually slightly expands, lowering the dark matter mass and increasing
the bias.

We decided not to include a stellar component or dark matter contraction to keep
our model simple. Moreover, when investigating the impact of the halo mass
determination on the inferred cosmological parameters, lower-mass haloes with
$m_\mathrm{500c} \lesssim 10^{14.5} \, \mh$ dominate the signal since they are
significantly more abundant and hence the fit is more sensitive to any bias in
this mass range. At low masses, all the aforementioned effects are clearly much
less important than the change in halo mass due to the missing halo gas. Hence,
we conclude that our model provides a reasonable estimate of the halo mass bias
induced by the change in halo density profiles due to the presence of baryons.

\section{Influence on cosmological parameter
  estimation}\label{sec:cosmo}

In this section we will investigate how the bias in the halo masses inferred
from mock weak lensing observations that we derived in
Section~\ref{sec:observations}, biases the measurement of cosmological
parameters from a number count analysis of a mock cluster sample.

\subsection{Mock cluster sample generation}\label{sec:cosmo:sample}

We create a cluster sample by drawing $(\log_{10}m_\mathrm{200m}, z)$ pairs from
the Poisson distribution with mean number density
\begin{align}
  \label{eq:dNdmdz}
  \frac{\diff N(m,z; \vect{\mathscr{C}})}{\diff \log_{10}m \, \diff z} & = \Omega_\mathrm{survey} \frac{\diff V_\mathrm{c}(z; \vect{\mathscr{C}})}{\diff z \diff \Omega} \frac{\diff n(m, z; \vect{\mathscr{C}})}{\diff \log_{10}m \, \diff z} \, ,
\end{align}
with the halo mass function $\diff n / \diff \log_{10}m \, \diff z$ of
\citet{Tinker2008} and the comoving volume $V_\mathrm{c}(z)$ for a
\citet{PlanckCollaboration2018g} cosmology with
$\vect{\mathscr{C}} \equiv \{\Omega_\mathrm{m}, \Omega_\mathrm{b},
\Omega_\Lambda, \sigma_8, n_\mathrm{s}, h\} = \{0.315, 0.049, 0.685, 0.811,
0.965, 0.674\}$. The sky area, $\Omega_\mathrm{survey}$, depends on the specific
survey. We use the \texttt{CCL}\footnote{\url{https://github.com/LSSTDESC/CCL}}
library to calculate the halo mass function \citep{Chisari2019}. We draw samples
from the non-homogeneous Poisson distribution by thinning the homogeneous
expectation on a grid of $(\log_{10}m_\mathrm{200m}, z)$ bins following the
method of \cite{Lewis1979}.

Since the \citet{Tinker2008} mass function was calibrated on DMO simulations,
the resulting mock cluster sample corresponds to a universe that contains only
dark matter. As we have shown in Section~\ref{sec:observations}, however, there
is a mismatch between the true halo mass, $m_\mathrm{200m,true}$, and the mass
of the equivalent DMO halo, $m_\mathrm{200m,dmo}$, due to the ejection of
baryons (see the middle panel of Fig.~\ref{fig:m200m_ratio_NFW}). Moreover, the
halo masses inferred from mock weak lensing observations, $m_\mathrm{200m,NFW}$,
can be biased with respect to the true halo mass (see the top panel of
Fig.~\ref{fig:m200m_ratio_NFW}). If these baryonic biases are not taken into
account in the cluster count analysis, the measured cosmological parameters will
be biased.

For each DMO halo in the cluster sample, we determine the biased halo mass
estimate of the corresponding halo with baryons,
$m_\mathrm{200m,NFW}(m_\mathrm{200m,dmo}, z)$, inferred from the NFW fits to the
mock weak lensing observations with either a fixed or free scale radius in
Section~\ref{sec:observations}. We interpolate the relation between the mass of
the halo including baryons and the mass of its equivalent DMO halo,
$m_\mathrm{500c}(m_\mathrm{200m,dmo}, z)$, from our halo model and determine the
corresponding mass ratio $m_\mathrm{200m,NFW} / m_\mathrm{200m,dmo}$ (see the
middle panel of Fig.~\ref{fig:m200m_ratio_NFW} for the ratio at $z=0.43$). We
will investigate how severely this baryonic mass bias affects the measured
cosmological parameters for stage III and stage IV-like surveys in
Sections~\ref{sec:cosmology_stageIII} and~\ref{sec:cosmology_stageIV},
respectively.

We start with a best-case scenario, where we have assumed a one-to-one mapping
between the observable mass proxy (e.g. the SZ detection significance) and the
halo masses inferred from weak lensing, i.e. we neglect the measurement
uncertainties in the mass estimation of individual clusters (we consider a more
realistic scenario in Sec.~\ref{sec:cosmology_stageIV}). This allows us to take
the weak lensing inferred halo masses as the starting point of our analysis.
When connecting haloes to their DMO equivalents, we also do not account for the
intrinsic scatter due to the differing mass distributions of individual haloes
that arise from their unique mass accretion histories. We assign the weak
lensing inferred halo masses to the DMO haloes without scatter. This is
consistent with our choice in Section~\ref{sec:hm:fits}, where we fit to the
median halo mass-binned cluster population of REXCESS, neglecting differences
between individual clusters in each mass bin.

Ignoring the mass estimation uncertainty and the intrinsic scatter in the halo
population would bias the observable--mass relation in an observed cluster
sample due to the preferential scatter of more abundant low-mass haloes into
higher mass bins. Hence, in a full cosmological analysis, converting the
observable to the true halo mass requires the inclusion of the mass estimation
uncertainty, and the intrinsic scatter in the halo population, while accounting
for the change in abundance of clusters as a function of mass and redshift. This
involves a joint fit to the abundance and the observable--mass relation of the
cluster sample as a function of cosmology \citep[see e.g.][]{Bocquet2019}. In
the more realistic scenario in Sec.~\ref{sec:cosmology_stageIV}, we will
implicitly assume that the scatter is constrained by the cluster abundances, so
that the precision of cosmological parameter estimation is not significantly
affected by not performing such a joint analysis.

\subsection{Stage III-like survey}\label{sec:cosmology_stageIII}

For a stage III-like cluster survey (e.g. SPT or DES; \citealt{Bocquet2019} and
\citealt{DESCollaboration2020}, respectively), we set the survey area to
$\Omega_\mathrm{survey}=2500 \, \mathrm{deg}^2$ to generate the cluster sample
using Eq.~\eqref{eq:dNdmdz}.

We want to quantify the statistical bias and uncertainty of the cosmological
parameters due to the baryonic halo mass bias. Hence, we generate 1000
independent cluster samples and fit the Maximum A-posteriori Probabilities (MAPs)
of the posterior distribution for each of the halo samples. We follow
\citet{Cash1979} and \citet{DeHaan2016} and obtain the posterior distribution for
the cosmological parameters
$\vect{\mathscr{C}} = \{\Omega_\mathrm{m}, \sigma_8, w_0\}$ by sampling the
Poisson likelihood, which is given up to a constant by
\begin{equation}
  \label{eq:ln_like}
  \ln \mathcal{L} \propto 2 \left( \sum_i \ln \frac{\diff N(m_i, z_i; \vect{\mathscr{C}})}{\diff m \, \diff z} - \int \diff m \, \diff z \, \frac{\diff N(m, z; \vect{\mathscr{C}})}{\diff m \, \diff z} \right) \, ,
\end{equation}
where $i$ runs over the individual clusters in the sample and the integral is
performed between $(z_\mathrm{min}=0.25, m_\mathrm{200m,min}=10^{14.5} \, \mh)$
and $(z_\mathrm{max}=2, m_\mathrm{200m,max}=10^{16} \, \mh)$. The lower bounds
are set by the sample selection and the upper bounds are chosen high enough that
the integral approaches the limit for $z, m \to \infty$. We assume flat prior
distributions $\Omega_\mathrm{m} \sim U(0.1, 0.6)$, $\sigma_8 \sim U(0.5, 1.1)$,
and $w_0 \sim U(-1.5, -0.5)$, where $U(a,b)$ indicates the uniform distribution
between $a$ and $b$. We fix the remaining cosmological parameters to the assumed
\citet{PlanckCollaboration2018g} values.

\begin{figure}
  \centering \includegraphics[width=\columnwidth]{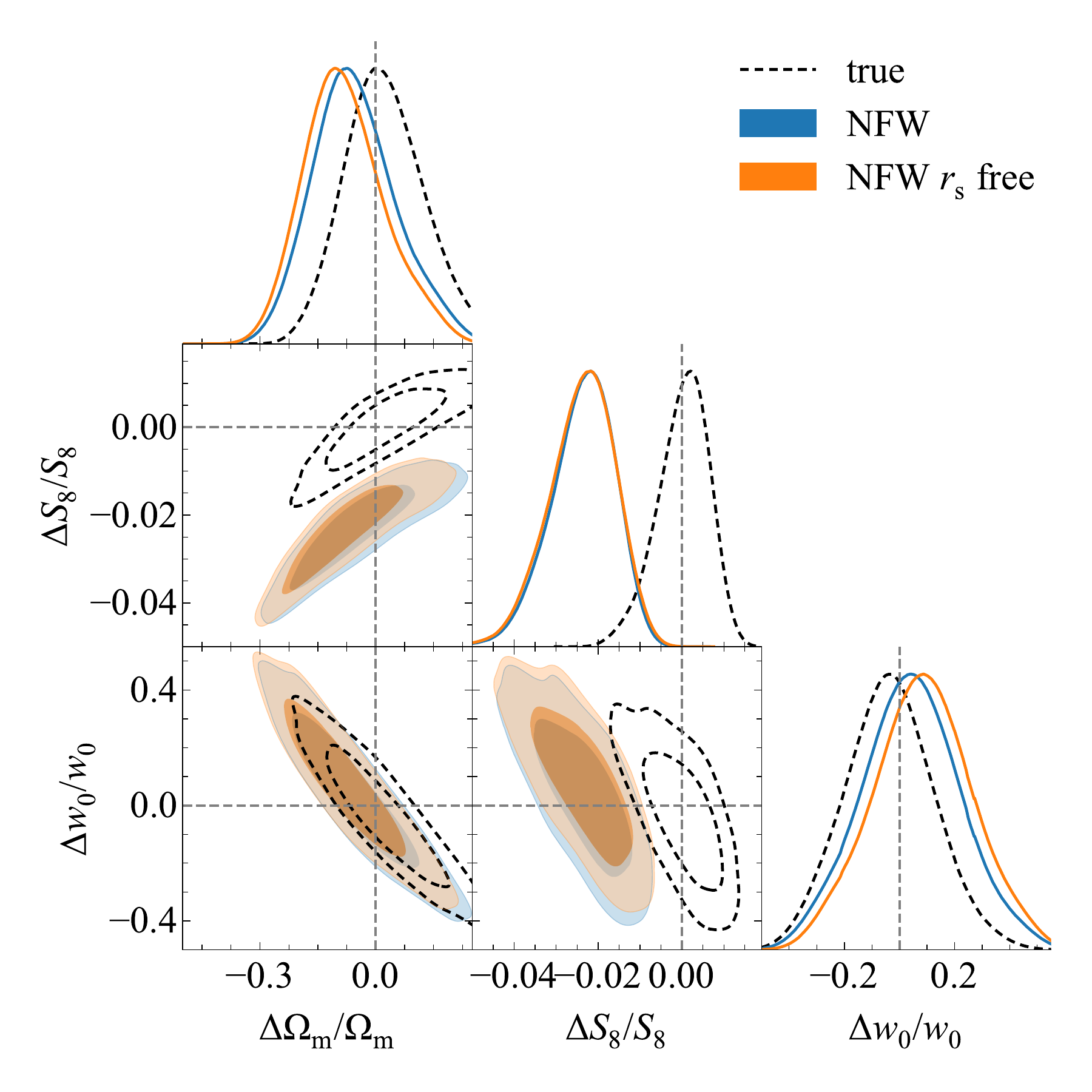}
  \caption{The distribution of the maximum a-posteriori probabilities in
    $(\Omega_\mathrm{m}, S_8=\sigma_8(\Omega_\mathrm{m}/0.3)^{0.2}, w_0)$ for
    1000 independent stage III-like cluster abundance surveys. Dashed contours
    show the results for a halo sample with no mass bias. Blue (orange) contours
    include a mass bias due to an NFW fit to mock weak lensing observations of
    the reduced shear, with a fixed (free) scale radius, $r_\mathrm{s}$. Neither
    $\Omega_\mathrm{m}$ nor $w_0$ are significantly biased due to baryonic
    effects. Relative constraints on $S_8$, however, are biased by \SWLIII\
    ($\approx 3 \, \sigma$) for both fixed and free scale radii in the NFW
    fit.}\label{fig:om_S8_w0_stageIII_maps}
\end{figure}

We show the resulting distribution of MAPs in
$(\Omega_\mathrm{m}, S_8=\sigma_8(\Omega_\mathrm{m}/0.3)^{0.2}, w_0)$ for each
of the different observational mass inferences in
Fig.~\ref{fig:om_S8_w0_stageIII_maps}. The dashed contours show the unbiased
halo sample. For this unbiased sample, all cosmological parameters are unbiased
and we find relative uncertainties of $\approx \pm 10 \, \percent$ in
$\Omega_\mathrm{m}$, $\approx \pm 0.7 \, \percent$ in $S_8$, and
$\approx \pm 16 \, \percent$ in $w_0$ for a current stage III-like cluster
survey. The quoted precision of all parameters underestimates the true
uncertainty, since we have performed an idealized analysis that does not include
observational uncertainties or intrinsic scatter in the derived halo masses, as
mentioned before. However, as we have already shown in
Fig.~\ref{fig:m200m_ratio_NFW}, the inferred halo masses are biased with respect
to the equivalent DMO halo mass due to the missing halo baryons. Hence, NFW
inferred halo masses with fixed and free scale radii (blue and orange contours,
respectively) are both predominantly biased in $S_8$, with a median bias and
16th-84th percentile uncertainties of $\Delta S_8/S_8 = \SWLIII$, where the
negative value indicates that $S_8$ is underestimated. Neither
$\Omega_\mathrm{m}$ nor $w_0$ show a significant bias for the different mass
determination methods. We list the cosmological parameter constraints for both
methods in Table~\ref{tab:cosmological_uncertainty_III}.

\begin{table*}
  \centering
  \caption{Inferred median bias and 16th-84th percentile statistical
    uncertainties of the individual best-fitting cosmological parameters for the
    different mass determination for a stage III-like survey with survey area
    $\Omega_\mathrm{survey}=2500 \, \mathrm{deg}^2$ and limiting redshift and
    halo mass
    $(z_\mathrm{min}=0.25, z_\mathrm{max}=2, m_\mathrm{200m,min}=10^{14.5} \,
    \mh)$. The columns correspond to cosmological parameters inferred from
    cluster samples with halo masses inferred from weak lensing fits with (a)
    fixed and (b) free NFW scale radii, and (c) the true cluster
    masses.}\label{tab:cosmological_uncertainty_III}
  \renewcommand{\arraystretch}{1.5}
  \begin{tabular}{l r r r}
    & $^{(\mathrm{a})}$NFW $r_\mathrm{s}$ fixed & $^{(\mathrm{b})}$NFW $r_\mathrm{s}$ free & $^{(\mathrm{c})}$true \\
    \hline
    \hline
    $\Delta \Omega_\mathrm{m} / \Omega_\mathrm{m}$ & \OmWLIII & \OmWLcIII & \OmtrueIII \\
    $\Delta S_8 / S_8$ & \SWLIII & \SWLcIII & \StrueIII \\
    $\Delta w_0 / w_0$ & \wWLIII & \wWLcIII & \wtrueIII \\
    \hline
  \end{tabular}
\end{table*}

The shifts in the cosmological parameters can be understood in the following
way. At a given redshift and for a fixed number count, the mass bias results in
an underestimation of the true halo mass. Hence, the number of clusters assigned
to the inferred halo mass is lower than it should be, since the number density
of clusters increases with decreasing mass. This underestimation is then
explained by decreasing the amount of structure in the Universe, assuming that
we are unaware of any mass bias.

In summary, current stage III-like cluster abundance surveys with ideal mass
estimations would find a biased cosmology (mainly in $S_8$) due to the mismatch
between $m_\mathrm{200m,true}$ and $m_\mathrm{200m,dmo}$. However, due to the
uncertainties induced by the mass estimation, which are larger than the
statistical uncertainty of our idealized survey, the baryonic mass bias is
currently not highly significant. As a reference, the current quoted
uncertainties for SPT (DES; \citealt{Bocquet2019} and
\citealt{DESCollaboration2020}, respectively) are $\pm 17 \, (17) \, \percent$
in $\Omega_\mathrm{m}$, $\pm 3 \, (6) \, \percent$ in $S_8$ (with $S_8$
definitions differing from ours for both SPT and DES), and
$\pm 26 \, (-) \, \percent$ in $w_0$ (DES does not constrain $w_0$),
respectively. These values exceed our statistical uncertainties of
$\pm 10 \, \percent$, $\pm 0.7 \, \percent$ and $\pm 16 \, \percent$,
respectively. The baryonic bias in the cosmological parameters that our model
predicts corresponds to a statistical significance of $0.5 \, \sigma$
($0.5 \, \sigma$) in $\Omega_\mathrm{m}$, $0.8 \, \sigma$ ($0.4 \, \sigma$) in
$S_8$, $0.3 \, \sigma$ ($-$) in $w_0$ for the precision of SPT (DES). In
Sec.~\ref{sec:cosmology_stageIV}, we show that the precision of the inferred
cosmological parameters is set by the accuracy with which the uncertainty in the
mass estimation is known. The mass estimation uncertainty is strongly degenerate
with $S_8$ and imposing an uninformative prior on the uncertainty of individual
cluster masses results in a significant decrease in the precision of the
constraint on $S_8$, in line with the comparison to SPT and DES.

\subsection{Stage IV-like survey}\label{sec:cosmology_stageIV}

\begin{figure}
  \centering \includegraphics[width=\columnwidth]{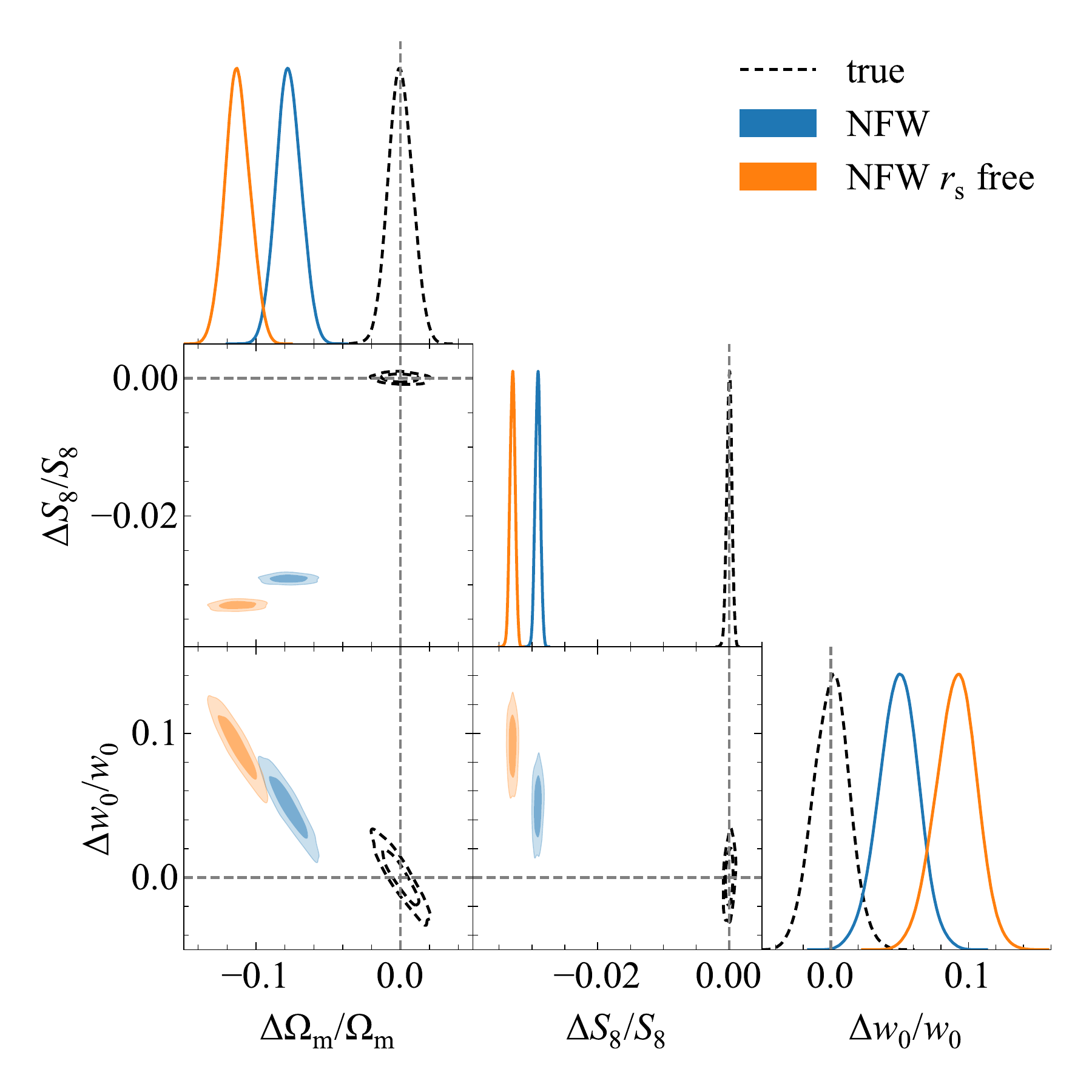}
  \caption{The distribution of the maximum a-posteriori probabilities in
    $(\Omega_\mathrm{m}, S_8=\sigma_8(\Omega_\mathrm{m}/0.3)^{0.2}, w_0)$ for
    1000 independent stage IV-like cluster abundance surveys. Dashed contours
    show the results for a halo sample with no mass bias. Blue (orange) contours
    include a mass bias due to an NFW fit to mock weak lensing observations of
    the reduced shear, with a fixed (free) scale radius, $r_\mathrm{s}$.
    Relative constraints on $S_8$ are very highly biased for both NFW fitting
    methods to the cluster density profiles including baryons. Similarly,
    $\Omega_\mathrm{m}$ and $w_0$ are biased by up to $13 \, \sigma$ and
    $6 \, \sigma$, respectively.}\label{fig:om_S8_w0_stageIV_maps}
\end{figure}

For a stage IV-like survey such as \emph{Euclid}, the survey area increases
dramatically to $\Omega_\mathrm{survey}=15000 \, \deg^2$. These surveys will
generally rely on observed galaxy overdensities to detect clusters and will,
consequently, have more complex selection functions that depend on the magnitude
limit of the survey \citep[see e.g.][]{Sartoris2016}. We take a simple mass cut
of $m_\mathrm{200m,min} = 10^{14} \, \mh$ and redshift cuts of
$z_\mathrm{min} = 0.1$ and $z_\mathrm{max} = 2$. Due to the increase in survey
area and the decrease in $m_\mathrm{200m,min}$, the total number of clusters
increases by about two orders of magnitude compared with a stage III-like
survey. The Poisson likelihood in Eq.~\eqref{eq:ln_like} becomes intractable,
especially if different mass calibrations are to be included, such as in
\citet{DeHaan2016}. We therefore switch to the Gaussian likelihood for bins
where the number of observed clusters, $N_\mathrm{obs}(m_i, z_j) > 10$
\begin{align}
  \label{eq:ln_like_gauss}
  \ln \mathcal{L} \propto \sum_{m_i, z_j} & -\frac{(N_\mathrm{obs}(m_i, z_j) - N(m_i, z_j; \mathscr{C}))^2}{2 N(m_i, z_j; \mathscr{C})} \\
  \nonumber
  & - \frac{\ln N(m_i, z_j; \mathscr{C})}{2} \, ,
\end{align}
and we use the Poisson likelihood for the other bins
\begin{align}
  \label{eq:ln_like_poisson_binned}
  \ln \mathcal{L} \propto \sum_{m_i, z_j} & N_\mathrm{obs}(m_i, z_j) \ln N(m_i, z_j; \mathscr{C}) - N(m_i, z_j; \mathscr{C}) \\
  \nonumber
                                          & - \ln N_\mathrm{obs}(m_i, z_j)! \, ,
\end{align}
where $(m_i, z_j)$ run over the logarithmic bins in $m_\mathrm{200m}$ and the
linear bins in $z$, respectively. We transition at the value
$N_\mathrm{obs}(m_i, z_j) = 10$ since Eq.~\eqref{eq:ln_like_gauss} is biased
with respect to Eq.~\eqref{eq:ln_like_poisson_binned} by a factor of
$1 + O(N_\mathrm{obs}^{-1/2})$, as worked out by \citet{Cash1979}. The Gaussian
likelihood makes it easier to include contributions from the sample variance,
which will also need to be included for the lower-mass haloes probed by stage
IV-like surveys \citep{Hu2003}. We have neglected the sample covariance in
generating our halo sample and, hence, we do not include it in our likelihood
analysis. We include the Poisson likelihood for the bins with low number counts
since the Gaussian likelihood cannot properly account for the discreteness of
the number count data, biasing the cosmological parameter estimates, as we show
in Appendix~\ref{app:mixed_likelihood}. In a more realistic setting, the sample
variance should be included in the cluster catalogue generation and the cluster
number count analysis. For stage IV-like surveys with low limiting masses, the
sample variance can dominate the shot noise, increasing the uncertainty on the
cluster number density, which reduces the bias for the bins with low number
counts. We choose 40 equally spaced bins between
$\log_{10} m_\mathrm{200m,min} /(\mh) = 14.0$ and the highest halo mass present
in each cluster sample. For the redshift, we take 8 equally spaced bins for
$z \in [0.1, 2]$. We assume the same priors as we did in
Section~\ref{sec:cosmology_stageIII}.

We show the resulting distribution of MAPs for the stage IV-like survey in
Fig.~\ref{fig:om_S8_w0_stageIV_maps}. The relative uncertainties for the
unbiased sample shrink to $\approx \pm 1.0 \, \percent$ in $\Omega_\mathrm{m}$,
$\approx \pm 0.04 \, \percent$ in $S_8$, and $\approx \pm 1.5 \, \percent$ in
$w_0$ for a stage IV-like cluster survey. Again, we stress that we underestimate
the true uncertainty, since we do not include any mass calibration
uncertainties. However, in our idealized analysis, the bias from ignoring
baryonic effects in the NFW inferred halo masses becomes catastrophic for $S_8$,
both for fixed and free scale radii. Moreover, we also find very significant
biases of up to $13 \, \sigma$ in $\Omega_\mathrm{m}$ and up to
$6 \, \sigma$ in $w_0$ (for the exact values, see
Table~\ref{tab:cosmological_uncertainty_IV}).

\begin{table*}
  \centering
  \caption{Inferred median bias and 16th-84th percentile statistical
    uncertainties of the individual best-fitting cosmological parameters for the
    different mass determination methods for a stage IV-like survey with
    $\Omega_\mathrm{survey}=15000 \, \mathrm{deg}^2$ and
    $(z_\mathrm{min}=0.1, z_\mathrm{max}=2, m_\mathrm{200m,min}=10^{14} \,
    \mh)$. The rows show cosmological parameters inferred from cluster samples
    with halo masses inferred from weak lensing fits with fixed or free NFW
    scale radii, and the true cluster masses. The columns show the results for
    samples with (a) ideal mass determinations, a mass uncertainty of
    $\pm 20 \, \percent$ either (b) marginalized over
    $\sigma_{\ln m} \sim N(\ln 1.2, \ln 1.02)$, or (c) included with a uniform
    prior $\sigma_{\ln m} \sim U(\ln 1.001, \ln 2)$ in the cosmological
    parameter estimation.}\label{tab:cosmological_uncertainty_IV}
  \renewcommand{\arraystretch}{1.5}
  \begin{tabular}{l l r r r}
    Mass uncertainty [\percent] & & $^{(\mathrm{a})}$ideal [$0$] & $^{(\mathrm{b})}$marg. [$\pm 20$] & $^{(\mathrm{c})}$uniform [$\pm 20$] \\
    \hline
    \hline
    NFW $r_\mathrm{s}$ fixed & $\Delta \Omega_\mathrm{m} / \Omega_\mathrm{m}$ & \OmWLIV & \OmWLmargtwentyIV & \OmWLscattertwentyIV \\
    & $\Delta S_8 / S_8$ & \SWLIV & \SWLmargtwentyIV & \SWLscattertwentyIV \\
    & $\Delta w_0 / w_0$ & \wWLIV & \wWLmargtwentyIV & \wWLscattertwentyIV \\
    \hline
    NFW $r_\mathrm{s}$ free & $\Delta \Omega_\mathrm{m} / \Omega_\mathrm{m}$ & \OmWLcIV & \OmWLcmargtwentyIV & \OmWLcscattertwentyIV \\
    & $\Delta S_8 / S_8$ & \SWLcIV & \SWLcmargtwentyIV & \SWLcscattertwentyIV \\
    & $\Delta w_0 / w_0$ & \wWLcIV & \wWLcmargtwentyIV & \wWLcscattertwentyIV \\
    \hline
    true & $\Delta \Omega_\mathrm{m} / \Omega_\mathrm{m}$ & \OmtrueIV & \OmtruemargtwentyIV & \OmtruescattertwentyIV \\
    & $\Delta S_8 / S_8$ & \StrueIV & \StruemargtwentyIV & \StruescattertwentyIV \\
    & $\Delta w_0 / w_0$ & \wtrueIV & \wtruemargtwentyIV & \wtruescattertwentyIV \\
    \hline
  \end{tabular}
\end{table*}

\begin{figure}
  \centering \includegraphics[width=\columnwidth]{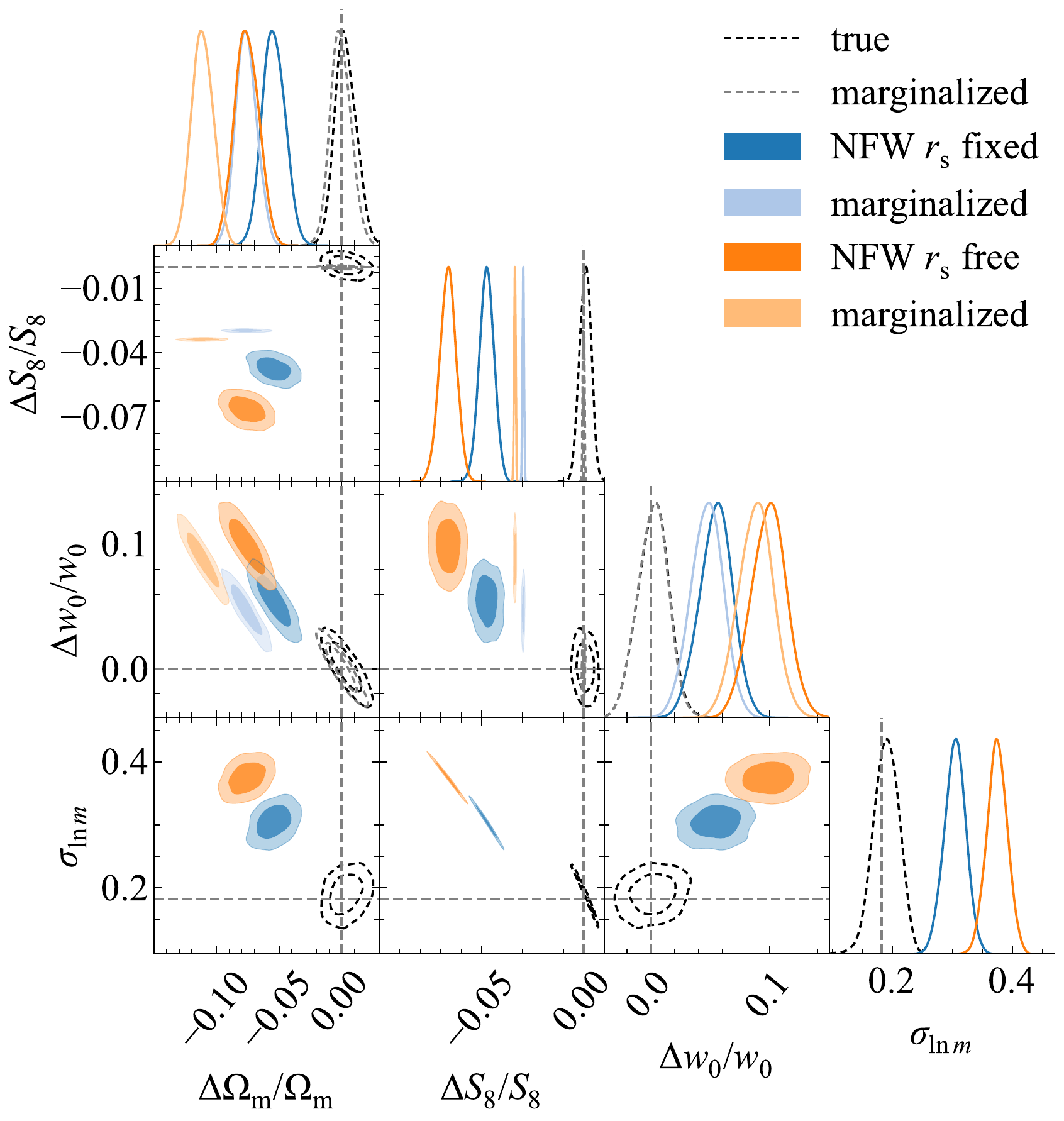}
  \caption{The marginalized maximum a-posteriori probability density functions
    for $\Omega_\mathrm{m}$, $S_8=\sigma_8(\Omega_\mathrm{m}/0.3)^{0.2}$, $w_0$,
    and $\sigma_{\ln m}$ for 1000 independent stage IV-like cluster abundance
    surveys with an uncertainty of $\pm 20 \, \percent$ on the individual
    cluster masses, assuming a mixed Gaussian-Poisson likelihood. Dashed
    contours show the results for a halo sample with no mass bias. Blue (orange)
    contours include a mass bias due to an NFW fit to mock weak lensing
    observations of the reduced shear, with a fixed (free) scale radius,
    $r_\mathrm{s}$. Dark contours also sample the individual cluster mass
    uncertainty $\sigma_{\ln m}$, whereas light contours have marginalized over
    a Gaussian distribution $\sigma_{\ln m} \sim N(\ln 1.2, \ln 1.02)$. Due to
    the preferential scatter of low-mass haloes into higher mass bins, an
    underestimate (overestimate) of $\sigma_{\ln m}$ for a fixed true value of
    $\sigma_{\ln m} = \ln 1.02$, results in an overestimate (underestimate) of
    $S_8$. Marginalizing over the mass uncertainty recovers the constraints
    obtained without mass uncertainty nearly
    identically.}\label{fig:cosmo_mass_uncertainty}
\end{figure}

However, the statistical precision of the cosmological parameters is overly
optimistic since we neglect any uncertainty on the individual cluster masses
inferred, resulting in extremely significant biases due to baryonic effects. For
stage IV-like surveys, the amplitude of the mean observable--mass relation can
reach \percent\ level accuracy due to the large number of clusters detected
\citep[see e.g.][]{Kohlinger2015}. However, the mass of an individual cluster
derived from the survey observable mass proxy will still have an uncertainty.
For an observable with a scatter of $\pm 20 \, \percent$ in the distribution
$P(m|\mathcal{O})$ of the true total halo mass, $m$, given the observable,
$\mathcal{O}$ \citep[similarly to the richness, see e.g.][]{Rozo2014, Mantz2016,
  Sereno2020a}, we expect an uncertainty of $\pm 20 \, \percent$ on the inferred
masses of an unbiased cluster sample.

In our idealized setting, we know the true underlying halo masses. We mimic the
uncertainty by adding a log-normal scatter of $\pm 20 \, \percent$ to the true
halo masses of the unbiased cluster samples and to the weak lensing inferred
halo masses of the biased cluster samples. We modify Eq.~\eqref{eq:dNdmdz} to
include an unknown mass uncertainty $\sigma_{\log_{10} m}$ for each mass bin
$i$, following \citet{Lima2005}
\begin{equation}
  \label{eq:dNdmdz_uncertainty}
  \frac{\diff N_{i}(m,z; \vect{\mathscr{C}})}{\diff \log_{10}m \, \diff z} = \frac{1}{2} \frac{\diff N(m,z; \vect{\mathscr{C}})}{\diff \log_{10}m \, \diff z}  (\mathrm{erfc}(x_i) - \mathrm{erfc}(x_{i+1})),
\end{equation}
where
\begin{equation}
  \label{eq:xi}
  x_i = \frac{\log_{10} m^\mathrm{obs}_{i} - \log_{10} m}{\sqrt{2 \sigma^2_{\log_{10} m}}} \, ,
\end{equation}
with $i$ and $i+1$ the edges of mass bin $i$. Adding the observational
uncertainty will result in haloes scattering to different mass bins, with each
bin gaining relatively more low-mass haloes due to their higher abundance. We
assume a uniform distribution for the mass uncertainty with
$\sigma_{\log_{10}m} \sim U(\log_{10}1.001, \log_{10} 2)$. In practice, we will
have some prior knowledge of the mass uncertainty of individual clusters. We
quantify this effect by including a cosmological analysis with a marginalization
over the mass uncertainty distribution
$\sigma_{\log_{10} m} \sim N(\log_{10} 1.2, \log_{10} 1.02)$, corresponding to
the case where the mass uncertainty is known to within $2 \, \percent$.

We show the resulting MAPs for the 1000 cluster samples with both an
uninformative prior (dark contours) and a marginalization (light contours) over
the individual cluster mass uncertainty in
Fig.~\ref{fig:cosmo_mass_uncertainty}. For the former case, we show the
posterior constraints on $\sigma_{\ln m}$ (which equals
$\sigma_{\log_{10} m} / \log_{10} e$, and approximately corresponds to the
\percent\ error on the halo mass). In the uninformative case, the mass
uncertainty is strongly degenerate with $S_8$, since an overestimate
(underestimate) of the true uncertainty would result in more (less) haloes
predicted to scatter into higher mass bins. At fixed observed number count
$N(m_{i}, z_{j})$, this effect is compensated by decreasing (increasing) $S_8$.

Compared to cluster samples with unbiased masses and no mass estimation
uncertainty, we find that the figure of merit (which we take as the inverse of
the area enclosed by $95 \, \percent$ of the surveys) for cluster samples with
unbiased masses and no prior knowledge of the cluster mass uncertainty of
$\pm 20 \, \percent$ (dark, dashed contours), decreases by factors of 7.1, 1.4,
and 7.6 in the $(\Omega_\mathrm{m}, S_8)$, $(\Omega_\mathrm{m}, w_0)$ and
$(S_8, w_0)$ planes, respectively. Similarly, the 1D marginalized regions
containing $68 \, \percent$ of the surveys increase by factors of 1.05, 7.7, and
1.02 for $\Omega_\mathrm{m}$, $S_8$, and $w_0$, respectively. However, with
accurate prior knowledge of the individual cluster mass estimation uncertainty
(light, dashed contours), the inferred cosmological parameters and their
precision are fully consistent with the ideal mass estimation case. This can be
seen by comparing the dashed and light dashed contours from
Figs.~\ref{fig:om_S8_w0_stageIV_maps} and~\ref{fig:cosmo_mass_uncertainty},
respectively, or by comparing the cosmological parameter constraints in columns
(a) and (b) for the true halo masses in
Table~\ref{tab:cosmological_uncertainty_IV}.

We find similar results when comparing the cluster samples that include a
baryonic bias and an uncertainty in the halo mass determination to samples that
include the baryonic bias but no mass estimation uncertainty. In the case of the
uniform prior on $\sigma_{\ln m}$ (dark, coloured contours), the figure of merit
in the $(\Omega_\mathrm{m}, S_8)$, $(\Omega_\mathrm{m}, w_0)$ and $(S_8, w_0)$
planes for a weak lensing fit with free (fixed) NFW scale radius, decreases by
factors of 11.2 (9.8), 1.7 (1.5), and 9.6 (8.9), respectively. Similarly, the 1D
marginalized regions for $\Omega_\mathrm{m}$, $S_8$, and $w_0$ containing
$68 \, \percent$ of the surveys, increase by factors of 1.1 (1.1), 10.1 (9.4),
and 1 (0.9), respectively. However, if the mass uncertainty is known to within
$2 \, \percent$ (light, coloured contours), then the ideal case is recovered
nearly identically. This can be seen by comparing the coloured and light
coloured contours from Figs.~\ref{fig:om_S8_w0_stageIV_maps}
and~\ref{fig:cosmo_mass_uncertainty}, respectively, or by comparing the
cosmological parameter constraints in columns (a) and (b) for the NFW fits with
fixed and free scale radii in Table~\ref{tab:cosmological_uncertainty_IV}. We
note that the distribution of the MAPs for the cluster samples with a baryonic
mass bias do not match between the informative and uninformative cases. This is
because the halo number counts calculated with Eq.~\eqref{eq:dNdmdz_uncertainty}
do not account for the mass-dependent baryonic bias. Hence, when leaving the
mass uncertainty as a free parameter, a more likely solution is found by
significantly increasing its value from the actual uncertainty, resulting in a
decrease in $S_8$. This does not happen for the cluster samples without a mass
bias.

\begin{figure}
  \centering \includegraphics[width=\columnwidth]{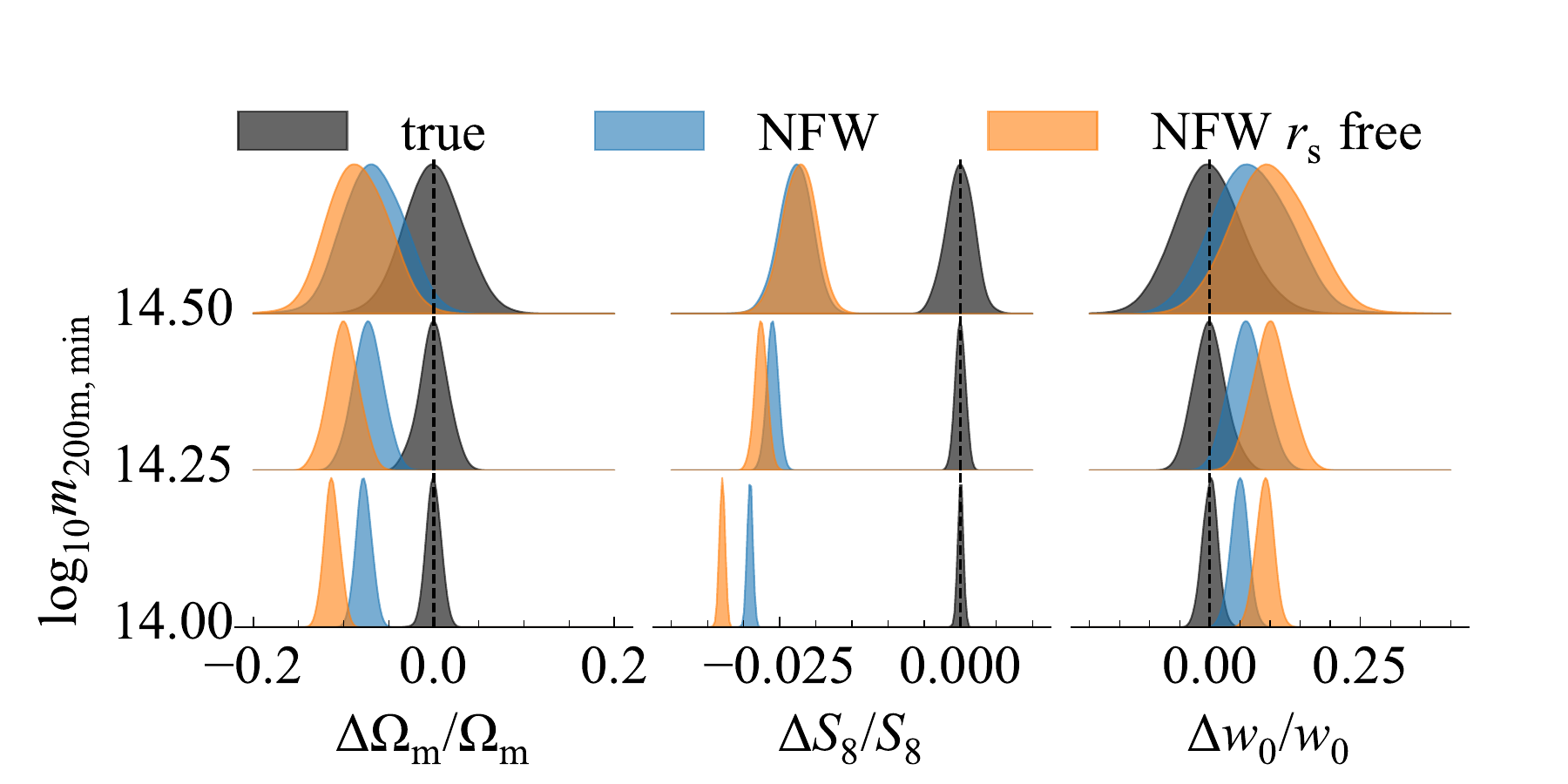}
  \caption{The marginalized maximum a-posteriori probability density functions
    for $\Omega_\mathrm{m}$, $S_8=\sigma_8(\Omega_\mathrm{m}/0.3)^{0.2}$ and
    $w_0$ for 1000 independent stage IV-like cluster abundance surveys with
    different mass cuts $m_\mathrm{200m,min}$ and perfect mass determinations.
    Gray PDFs show the results for a halo sample with no mass bias. Blue
    (orange) PDFs include a mass bias due to baryonic effects resulting from an
    NFW fit to mock weak lensing observations of the reduced shear, with a fixed
    (free) scale radius, $r_\mathrm{s}$. The bias in $S_8$ is reduced by a
    factor of $\approx 8$ if the mass cut is increased from
    $m_\mathrm{200m,min} = 10^{14} \, \mh$ to
    $m_\mathrm{200m,min} = 10^{14.5} \, \mh$, but is still highly significant,
    while the bias in $\Omega_\mathrm{m}$ and $w_0$ is reduced to within
    $2.5 \sigma$.}\label{fig:cosmo_1d_mcut}
\end{figure}

Any bias in the cosmological parameters can be reduced at the expense of a
larger uncertainty by increasing the mass cut of the cluster sample. We show the
marginalized 1D probability density functions for the cosmological parameters
for cluster samples without mass uncertainties using different limiting masses
in Fig.~\ref{fig:cosmo_1d_mcut}. Increasing the mass cut from
$m_\mathrm{200m,min}=10^{14} \, \mh$ to $m_\mathrm{200m,min}=10^{14.5} \, \mh$
reduces the bias in $S_8$ by a factor $\approx 8$ to $10 \, \sigma$, while the
bias in $\Omega_\mathrm{m}$ and $w_0$ is reduced to within $2.5 \, \sigma$.
However, this increase in the mass cut comes at the expense of a large increase
of the statistical errors.

In reality, there will be extra uncertainties due to the photometric redshift
estimation of the clusters and the lensed source galaxies, which will scatter
clusters between redshift and mass bins. Moreover, the mass uncertainty is a
combination of observational systematic uncertainties that evolve differently
with mass and redshift \citep{Kohlinger2015}. We have shown that the precision
of the inferred cosmological parameters will ultimately be set by the accuracy
with which the mass uncertainty of individual cluster masses can be determined.
The accuracy of the inferred cosmological parameters will depend on how
accurately the bias between the inferred halo masses and the equivalent DMO halo
masses can be determined.

Our results clearly indicate the need for more advanced mass inference methods
from weak lensing observations and a better calibration between the observed and
theoretical halo masses. Under our assumption that the dark matter distribution
is not significantly affected by the presence of baryons, it is possible to
obtain unbiased halo mass estimates. This suggests that combining measurements
of the total and baryonic halo mass, through, e.g., combined weak lensing and
X-ray or SZ observations, respectively, would provide significantly less biased
mass estimates of the dark matter mass and hence, after scaling by the universal
baryon fraction, of the equivalent DMO halo. In
Section~\ref{sec:aperture_masses}, we explore the possibility of using aperture
masses, which are less sensitive to the assumed halo density profile.

\section{Aperture masses}\label{sec:aperture_masses}

In Section~\ref{sec:observations}, we found that we cannot infer unbiased
equivalent DMO halo masses from mock weak lensing observations, even when the
inferred total halo mass is unbiased. This follows from the deviation of the
baryonic component from the assumed NFW density profile and the fact that the
baryon fraction is smaller than the universal value in the radial range of the
weak lensing observations.

It might be necessary to rethink how we link observed haloes to the theoretical
halo mass function, since this is the main baryonic uncertainty. Preferably, the
inferred halo masses should differ as little as possible from their equivalent
DMO haloes. It has been shown by \citet{Herbonnet2020} that projected halo
masses derived from a weak lensing analysis capture the true projected halo mass
more accurately than deprojected methods can. The aperture mass is a powerful
tool, because it can be computed directly from the data under minimal
assumptions about the halo density profile \citep[see e.g.][]{Clowe1998,
  Hoekstra2015}. Moreover, we would expect the mass enclosed in a sufficiently
large aperture to converge to the equivalent DMO halo mass as long as a larger
fraction of the cosmic baryons is included for a larger aperture.

\begin{figure}
  \centering
  \includegraphics[width=\columnwidth]{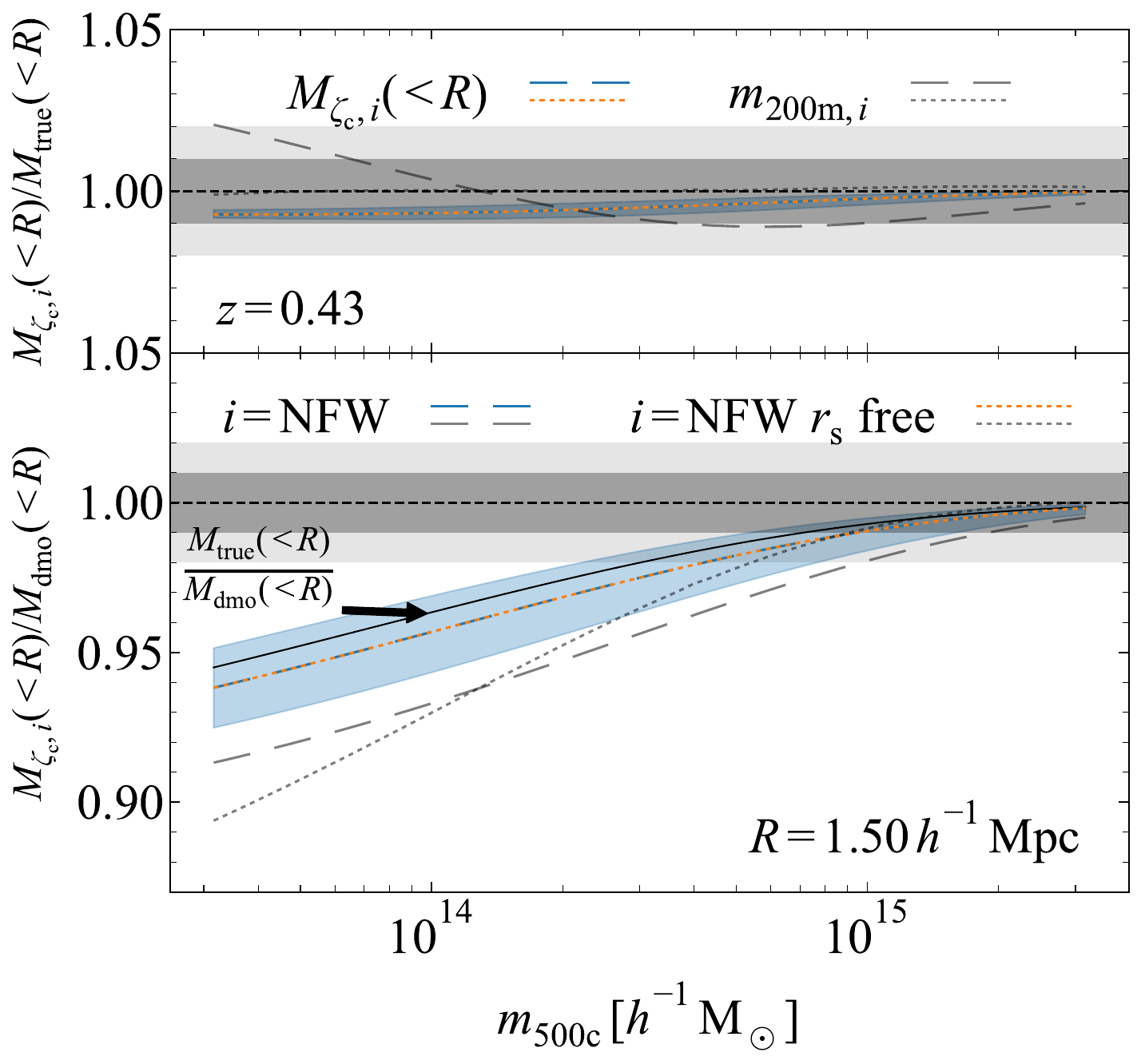}
  \caption{\emph{Top panel:} The ratio of the total aperture mass within
    $R < 1.5 \, \mpch$ derived from mock weak lensing observations, to the true
    aperture mass. The coloured, dashed and dotted lines show the ratio of the
    aperture masses inferred for the best-fitting NFW density profiles, with
    fixed and free scale radius, respectively, to the true aperture mass using
    the statistic from Eq.~\eqref{eq:M_ap}. The gray dashed and dotted lines
    show the ratio of the measured and true deprojected masses,
    $m_\mathrm{200m,NFW}/m_\mathrm{200m,true}$, for the same NFW fits. For the
    aperture masses, there is practically no difference between using a fixed or
    free NFW scale radius, indicating insensitivity to the assumed density
    profile. The derived aperture mass is within $1 \, \percent$ of the true
    aperture mass for all halo masses. \emph{Bottom panel:} The ratio of the
    total aperture mass within $R < 1.5 \, \mpch$ derived from mock weak lensing
    observations, to the same aperture mass of the equivalent DMO halo. Line
    styles are the same as in the top panel. The ratio of the true to the
    equivalent DMO halo aperture mass is shown as the solid, black line. The
    aperture masses are less biased with respect to the equivalent DMO mass than
    the deprojected masses, $m_\mathrm{200m}$, which are shown as the gray
    lines. }\label{fig:M_ap_vs_dmo}
\end{figure}

We have performed aperture mass measurements of our mock weak lensing data in
the following way. First, we convert the reduced shear to the tangential shear,
assuming the best-fitting NFW density profile with a fixed or free scale radius,
to compute $\kappa_\mathrm{NFW}(R)$
\begin{equation}
  \label{eq:gamma_T_ap}
  \gamma_\mathrm{T}(R) = (1 - \kappa_\mathrm{NFW}(R)) g_\mathrm{T}(R) \, .
\end{equation}
Here, the difference between $\kappa_\mathrm{NFW}(R)$ and the true convergence is
$\lesssim 2 \, \percent$ over the radial range of the observations, resulting in
negligible error due to the wrong density profile assumption. Then, we compute
the aperture mass using the statistic introduced by \citet{Clowe1998}
\begin{align}
  \nonumber
  \zeta_\mathrm{c}(R_1) & = \mean{\kappa}(R{<}R_1) - \mean{\kappa}(R_2 < R \leq R_\mathrm{max})  \\
  \label{eq:zeta_clowe}
                        & = 2 \int\limits_{R_1}^{R_2} \amean{\gamma_\mathrm{T}} \diff \ln R + \frac{2}{1 - R_2^2/R_\mathrm{max}^2} \int\limits_{R_2}^{R_\mathrm{max}} \amean{\gamma_\mathrm{T}} \diff \ln R \, ,
\end{align}
where $\amean{\gamma_\mathrm{T}}$ is the azimuthally averaged tangential shear,
for which we use the tangential shear from Eq.~\eqref{eq:gamma_T_ap}, derived
from mock weak lensing observations of the reduced shear. The aperture mass is
then given by
\begin{equation}
  \label{eq:M_ap}
  M(R{<}R_1) = \pi R^2 \Sigma_\mathrm{crit}(\zeta_\mathrm{c}(R_1) + \mean{\kappa}(R_2 < R \leq R_\mathrm{max})) \, ,
\end{equation}
where we can use the best-fitting NFW profile to determine
$\mean{\kappa}(R_2 < R \leq R_\mathrm{max})$, which is a small correction that
again differs negligibly from the true convergence profile due to the NFW
assumption. The aperture masses inferred from the above equations recover the
true projected halo mass at sub-\percent\ accuracy over the entire mass range,
as we show in the top panel of Fig.~\ref{fig:M_ap_vs_dmo}. Aperture masses are
thus a very accurate measure of the true enclosed halo mass, more so due to the
fact that they depend so little on assumptions about the underlying true density
profile.

However, the problem of linking the observed haloes to their equivalent DMO
counterparts still remains, although it is slightly alleviated. In the bottom
panel of Fig.~\ref{fig:M_ap_vs_dmo}, we show the ratio of the aperture masses
from mock weak lensing observations within a fixed aperture of
$R < 1.5 \, \mpch$ to the mass of the equivalent DMO halo within the same
aperture. We choose this aperture size since it is within the range of our mock
weak lensing observations in Section~\ref{sec:observations} and it is larger
than the fixed overdensity radius $r_\mathrm{200m}$ for haloes with
$m_\mathrm{500c} < 10^{14.5} \, \mh$, for which
$r_\mathrm{200m}(z=0.43) \approx 1.3 \, \mpch$, resulting in a larger fraction
of the universal baryons within it for these abundant haloes. We choose the
outer annulus for the correction factor in Eq.~\eqref{eq:zeta_clowe} between
$R_2=2.4 \, \mpch$ and $R_\mathrm{max}=2.5 \, \mpch$ such that the NFW
correction term is small compared to $\zeta_\mathrm{c}$. Aperture masses perform
better at recovering the mass of the linked DMO halo than the deprojected NFW
masses in Sec.~\ref{sec:observations} as long as $R_1 \gtrsim r_\mathrm{200m}$,
i.e. for all haloes with $m_\mathrm{500c} \lesssim 4 \times 10^{14} \, \mh$ at
$z=0.43$, as can be seen from the comparison of the coloured dashed and dotted
lines with the gray lines in the bottom panel of Fig.~\ref{fig:M_ap_vs_dmo}.
This follows from the fact that the halo baryon fractions converge to the cosmic
value in the cluster outskirts. This is one of our main conclusions: to link
observed haloes to their DMO equivalents, we need to make sure that we are
accounting for the ejected baryons. Otherwise, any mass estimate, while not
necessarily biased with respect to the \emph{true halo} mass, will be biased
with respect to the \emph{equivalent DMO halo} mass. It is this latter bias that
is fatal for accurate cluster cosmology.

The fact that the statistic in Eq.~\eqref{eq:M_ap} is practically unbiased with
respect to the true aperture mass, regardless of the assumed density profile,
makes it an appealing alternative to the deprojected mass determination methods.
The problem of calibrating the observed halo masses to their equivalent DMO
counterparts, while alleviated, still remains. Since there are so far no
theoretical calibrations for the halo aperture mass function, we do not check
the performance of the aperture mass determinations for cluster cosmology.

\section{Discussion}\label{sec:discussion}

We have introduced a phenomenological model that reproduces the baryon content
inferred from the X-ray surface brightness profiles of the average observed
cluster population in the REXCESS survey. We have shown how we can include
observed baryonic density profiles in a halo model, while ensuring that the halo
baryon fraction converges to the cosmic value in the halo outskirts, by fitting
the inferred radial halo baryon fraction with the correct asymptotic value. By
assuming that baryons do not significantly alter the distribution of the dark
matter, we were able to link observed haloes to their equivalent haloes in a DMO
universe, which allowed us to predict their number density. Then, we performed
mock weak lensing observations to quantify the effect of the changing halo
density profile due to the ejection of baryons on the inferred halo masses.
Finally, we investigated the bias due to baryons in the measured cosmological
parameters from a number count analysis of a mock cluster sample with masses
inferred from weak lensing observations. We have justified that our
simplifications result in robust lower bounds on the amplitude of the shift due
to baryons of both cluster masses and cosmological parameters from an idealized
cluster count cosmology analysis. The survey-specific systematic uncertainties
set the statistical significance of these shifts. We have shown that the
baryonic bias in the cosmological parameters is highly significant even when not
including prior knowledge of the uncertainty in the cluster mass inferred from
an observable mass proxy. Now we situate our results in the wider context of the
literature.

\citet{Balaguera-Antolinez2013} studied the effect of baryons on the
cosmological parameters inferred from cluster counts. They used the observed
baryon fractions of clusters to infer their equivalent DMO halo masses,
similarly to our method. They also find significant biases in the inferred
cosmological parameters, mainly a strong suppression in $\Omega_\mathrm{m}$ and
a slight increase in $\sigma_8$ (the exact bias in $\sigma_8$ depends on their
chosen cluster baryon fraction relation). The amplitude and direction of the
bias differ from ours as \citet{Balaguera-Antolinez2013} use a single, smaller
mock cluster sample ($\approx 2.8 \times 10^4$ clusters compared to
$\approx 1.7 \times 10^5$) that spans a lower redshift range and they did not
include the effect of baryons on the cluster weak lensing mass determinations.

Previously, weak lensing mass determinations have been studied in both DMO
\citep[e.g.][]{Bahe2012} and hydrodynamical simulations
\citep[e.g.][]{Henson2017, Lee2018a}. While \citet{Bahe2012} and
\citet{Henson2017} find different values for the mass bias, i.e.
$\approx 5 \, \percent$ and $\approx 10 \, \percent$, respectively, they both
conclude that these biases result from fitting complex, asymmetric clusters with
idealized NFW profiles. (Importantly, these analyses leave the concentration
free, which is not the case in most observational analyses.) If this is the
case, then we could reduce the weak lensing mass bias by performing a stacked
analysis, if we have an unbiased cluster sample. Or, since \citet{Henson2017}
find a similar bias at fixed halo mass for haloes in both hydrodynamical and DMO
simulations (see the top panel of their fig. 11), it seems feasible to model the
mass bias due to triaxiality, substructures and departures from the NFW shape,
by performing mock observations of DMO haloes \citep[as in
e.g.][]{Dietrich2019a}. However, we have shown, as has also been pointed out by
\citet{Lee2018a}, that the distribution of observed cluster baryons implies an
intrinsic difference in the density profiles between observed clusters and their
DMO equivalents that cannot be captured when assuming a fixed
concentration--mass relation. Hence, the inferred halo masses would still be
biased, even when accounting for the bias due to halo asymmetry. We found that
leaving the concentration of the haloes free mitigates this baryonic mass bias,
as was also shown in \citet{Lee2018a}. However, we showed that the bias in the
measured cosmological parameters from a cluster count analysis actually
\emph{increases} when leaving the concentration--mass relation free in the weak
lensing analysis. This is because low-mass cluster masses are overestimated when
fixing the concentration--mass relation, which compensates for some of the
missing baryons and thus reduces the bias with respect to the equivalent DMO
halo mass for these abundant clusters.

For cluster cosmology, the vital part is then linking these inferred cluster
masses to the equivalent DMO haloes whose number counts we can predict, as
argued by \citet{Cui2014}, \citet{Cusworth2014} and \citet{Velliscig2014}. In
the cosmo-\textsc{OWLS} simulations, \citet{Velliscig2014} found differences of
$\lesssim 1 \, \percent$ between cluster masses in the hydrodynamical and DMO
simulations for clusters with $m_\mathrm{500c} \gtrsim 10^{14.5} \, \mh$. In our
model, we only find such small biases for haloes with masses
$m_\mathrm{500c} \gtrsim 10^{15} \, \mh$. As discussed previously, if we
optimistically assume that the predictions from cosmo-\textsc{OWLS} are correct,
then this difference could be due to our neglect of the back-reaction of the
baryons on the dark matter, and the stellar component. However, for low-mass
haloes ($m_\mathrm{500c} \lesssim 10^{14.5} \, \mh$), which will dominate the
signal for stage IV-like surveys, these effects are negligible compared to the
mass suppression due to the missing baryons.

Using the Magneticum simulation set, \citet{Bocquet2016} and \citet{Castro2020}
studied the change in the halo mass function due to baryons and its impact on
cluster cosmology. \citet{Bocquet2016} performed a cluster count analysis using
halo mass functions calibrated on DMO simulations, to measure the cosmological
parameters from a cluster sample generated from the halo mass function of their
hydrodynamical simulation. They did not find significant biases for stage
III-like surveys, but their shifts in $\Omega_\mathrm{m}$ and $S_8$ for an
eROSITA-like survey are qualitatively similar to our stage IV-like survey
predictions. The difference for the stage III-like surveys could be caused by a
smaller mismatch between the halo masses in their hydrodynamical and DMO
simulations than we infer from observations.

\citet{Castro2020} made Fisher forecasts for a joint cluster number count and
clustering analysis of a \emph{Euclid}-like survey using the baryonic and DMO
halo mass functions in the Magneticum simulations. They confirmed that
correcting for the baryonic mass bias brings the different halo mass functions
into closer agreement. However, they find less significant baryonic mass
suppression than we do. The resulting biases in the cosmological parameters are
significantly smaller than what we find. This difference is most likely caused
by both the lower baryonic mass suppression in Magneticum and a different sample
selection. We have chosen a minimum redshift $z_\mathrm{min} = 0.1$ and a
constant limiting mass cut of $m_\mathrm{200m,min} = 10^{14} \, \mh$, whereas
\citet{Castro2020} use $z_\mathrm{min} = 0.2$ and a redshift-dependent mass
threshold varying around $m_\mathrm{200c,min} \approx 10^{14} \, \mh$ within
$0.1 \, \mathrm{dex}$ (see their fig.~13). Consequently, our sample includes
more low-mass clusters which increases the statistical significance of the bias
(as we show in Fig.~\ref{fig:cosmo_1d_mcut}).

An important difference between our work and previous work is that we have used
a phenomenological model that reproduces the \emph{observed} baryon content of
clusters. Hence, we do not suffer from the uncertainty related to the assumed
subgrid models in hydrodynamical simulations. We only rely on the fact that
hydrodynamical simulations imply that the baryonic mass suppression of matched
haloes explains the difference between their halo mass function and that derived
from DMO simulations. All in all, even though the exact value of the baryonic
mass bias between observed and equivalent DMO halo masses, and, consequently,
the halo mass function, can differ by up to a few \percent\ depending on which
simulations or observations are used, the general behaviour is the same and
implies the need to account for baryonic effects in cluster count cosmology.

\section{Conclusions}\label{sec:conclusions}

We set out to investigate the implications for cluster count cosmology of the
disconnect between the robust theoretical understanding of cluster-sized
($m_\mathrm{500c} > 10^{14} \, \mh$) dark matter-only haloes and the observed
cluster population, an issue which was pointed out by \citet{Cui2014},
\citet{Cusworth2014}, and \cite{Velliscig2014}. They found that in
hydrodynamical simulations, there is a significant mismatch between the enclosed
halo masses at fixed radius that is determined by the halo baryon fraction. We
study how the change in the halo density profiles due to the observed
distribution of baryons affects the estimated masses from mock weak lensing
observations and the resulting cosmological parameters from a cluster number
count analysis.

Our model relies on X-ray observations from the REXCESS data \citep{Croston2008}
to constrain the baryonic density profile of cluster-mass haloes. Under the
assumption that the dark matter density profile does not change significantly in
the presence of baryons, we can link observed haloes to their DMO equivalents.
The distribution of a fraction of the DMO halo mass, i.e. the cosmic baryon
fraction, will change in the observed halo. Once this link has been established,
we can study the change resulting from this baryonic mass bias in cosmological
parameters inferred from a number count analysis. We showed that the currently
standard weak lensing mass calibrations that assume NFW density profiles and a
fixed concentration--mass relation from DMO simulations, are inherently biased
for cluster-mass haloes. Fixing the concentration of the halo results in
underestimated halo masses since baryons are ejected beyond the typical radial
range that the weak lensing observations are sensitive to. The density profile
is fit out to radii where baryons are missing and is not flexible enough to
capture the increase in baryonic mass towards larger radii. However, we showed
that there is enough freedom in the NFW density profile to provide unbiased halo
mass estimates if the concentration is left free (see
Fig.~\ref{fig:mr_NFW_vs_true}), in agreement with \citet{Lee2018a}.

However, even unbiased total halo masses result in biased cosmological parameter
estimates because of the mismatch between the observed haloes and their DMO
equivalents due to ejected baryons (see the middle panel of
Fig.~\ref{fig:m200m_ratio_NFW}). This is the dominant baryonic bias. A fiducial
weak lensing analysis with fixed concentration--mass relation for a stage
IV-like survey would result in highly significantly biased estimates of the
cosmological parameters, underestimating $\Omega_\mathrm{m}$ and $S_8$ by up to
$9 \, \sigma$ and $76 \, \sigma$, respectively, and overestimating $w_0$ by
$3.5 \, \sigma$ (see Fig.~\ref{fig:om_S8_w0_stageIV_maps} and
Table~\ref{tab:cosmological_uncertainty_IV} for the exact values of the bias).
Although leaving the concentration--mass relation free in the weak lensing
analysis decreases the bias in the total mass, it actually \emph{increases} the
bias in the cosmological parameters to $13 \, \sigma$, $82 \, \sigma$ and
$6 \, \sigma$, respectively. This is because the masses of low-mass clusters are
overestimated when fixing the concentration--mass relation, which results in a
smaller bias compared to the equivalent DMO mass.

We showed that including a constant uncertainty of $\pm 20 \, \percent$ in the
individual, unbiased cluster masses only reduces the precision of the inferred
cosmological parameters if the mass uncertainty itself is not accurately
determined. An uninformative prior on the mass uncertainty decreases the
precision of $\Omega_\mathrm{m}$, $S_8$, and $w_0$ by factors of 1.05, 7.8, and
1.02, respectively. However, assuming the mass uncertainty of individual
clusters is known to within $\pm 2 \, \percent$ results in constraints that are
nearly identical to those derived from ideal cluster masses (see
Fig.~\ref{fig:cosmo_mass_uncertainty} and
Table~\ref{tab:cosmological_uncertainty_IV}).

The picture changes slightly for cluster samples that include the baryonic
mass bias. To quantify how neglecting the baryonic mass bias affects the
inferred cosmological parameters, we do not account for the mass-dependent
baryonic bias when fitting the cluster number counts. Since the model without
prior knowledge of the mass uncertainty can vary the mass uncertainty as well
as the cosmological parameters, the best-fitting parameters differ between the
cases with and without prior knowledge of the mass uncertainty. The
uninformative prior on the mass uncertainty decreases the precision of
$\Omega_\mathrm{m}$, $S_8$, and $w_0$ by factors of up to 1.1, 10.7, and 1.02,
respectively. Knowing the mass uncertainty to within $\pm 2 \, \percent$ again
results in constraints that cannot be distinguished from those derived from
ideal cluster masses (see Fig.~\ref{fig:cosmo_mass_uncertainty} and
Table~\ref{tab:cosmological_uncertainty_IV}). The baryonic bias is thus highly
statistically significant, even in the presence of mass estimation
uncertainties. The accuracy of the cosmological parameters inferred from
cluster number counts depends on how accurately inferred halo masses can be
linked to their equivalent DMO halo masses. The precision of the cosmological
parameter estimates is determined by how accurately the individual cluster
mass estimation uncertainty is known.

For stage III-like surveys and assuming a fixed (free) concentration--mass
relation, we found biases of $\approx 0.6 \, \sigma$ ($0.9 \, \sigma$),
$3 \, \sigma$ ($3 \, \sigma$) and $0.1 \, \sigma$ ($0.5 \, \sigma$) in
$\Omega_\mathrm{m}$, $S_8$, and $w_0$, respectively, again, assuming ideal
cluster mass estimations (see Fig.~\ref{fig:om_S8_w0_stageIII_maps} and
Table~\ref{tab:cosmological_uncertainty_III}). However, we stressed that the
uncertainties induced by the mass estimation for current stage III-like surveys
exceed the statistical uncertainty of our idealized survey.

We also measured aperture masses, since they are expected to provide less biased
estimates of the total projected mass than deprojected mass estimates,
independently of the assumed density profile of the cluster (see the top panel
of Fig.~\ref{fig:M_ap_vs_dmo}) and they are more closely related to the actual
weak lensing observable \citep[e.g.][]{Clowe1998, Herbonnet2020}. However, even
though it is slightly alleviated, the problem of linking observed haloes to
their DMO equivalents remains (see the middle panel of
Fig.~\ref{fig:M_ap_vs_dmo}). We expect the total projected mass to approach the
projected DMO mass at large radii \citep{VanDaalen2014a}. One problem is that
correlated large-scale structure becomes important near the cluster virial
radius \citep[e.g.][]{Oguri2011a}, which requires accurate modelling of the
cluster-mass halo bias. We did not include this effect in our model. Using
aperture mass estimates would also require a recalibration of halo mass function
predictions to this observable.

Any attempt to use clusters for cosmology will need to include a robust method
for linking observed haloes to their DMO equivalents. A joint approach,
combining weak lensing observations with, for example, hot gas density profiles
from from X-ray telescopes like eROSITA---and, in the future, Athena---and/or SZ
observations would allow the reconstruction of the cluster dark matter mass,
which has already been shown to be much less biased with respect to the
equivalent DMO halo mass \citep{Velliscig2014}. This is an essential avenue to
be explored. If we cannot robustly establish the link to DMO haloes, we cannot
obtain unbiased cosmological parameters.

\section*{Acknowledgements}

We would like to thank the referee for a constructive report that helped us
clarify the structure and the implications of our findings. SD would like to
thank Marius Cautun for useful discussions about likelihoods. This work is part
of the research programme Athena with project number 184.034.002 and Vici grants
639.043.409 and 639.043.512, which are financed by the Dutch Research Council
(NWO).

\section*{Data availability}

The 1000 mock cluster samples for the stage III-like cluster survey and the MAPs
for both the stage III and stage IV-like surveys are publically available
through Zenodo at
\href{https://zenodo.org/record/4469436}{10.5281/zenodo.4469436}. The stage
IV-like mock cluster samples can be obtained upon request since they exceed the
file size limit of Zenodo. The code for the analysis is available at
\url{https://github.com/StijnDebackere/lensing_haloes/}.

%%%%%%%%%%%%%%%%%%%%%%%%%%%%%%%%%%%%%%%%%%%%%%%%%% 


\bibliographystyle{mnras}
\bibliography{halo_masses.bbl}

%%%%%%%%%%%%%%%%%%%%%%%%%%%%%%%%%%%%%%%%%%%%%%%%%%

%%%%%%%%%%%%%%%%%%%% APPENDICES %%%%%%%%%%%%%%%%%%

\appendix

\section{Model fits}\label{app:model_fits}

\begin{figure}
  \centering \includegraphics[width=\columnwidth]{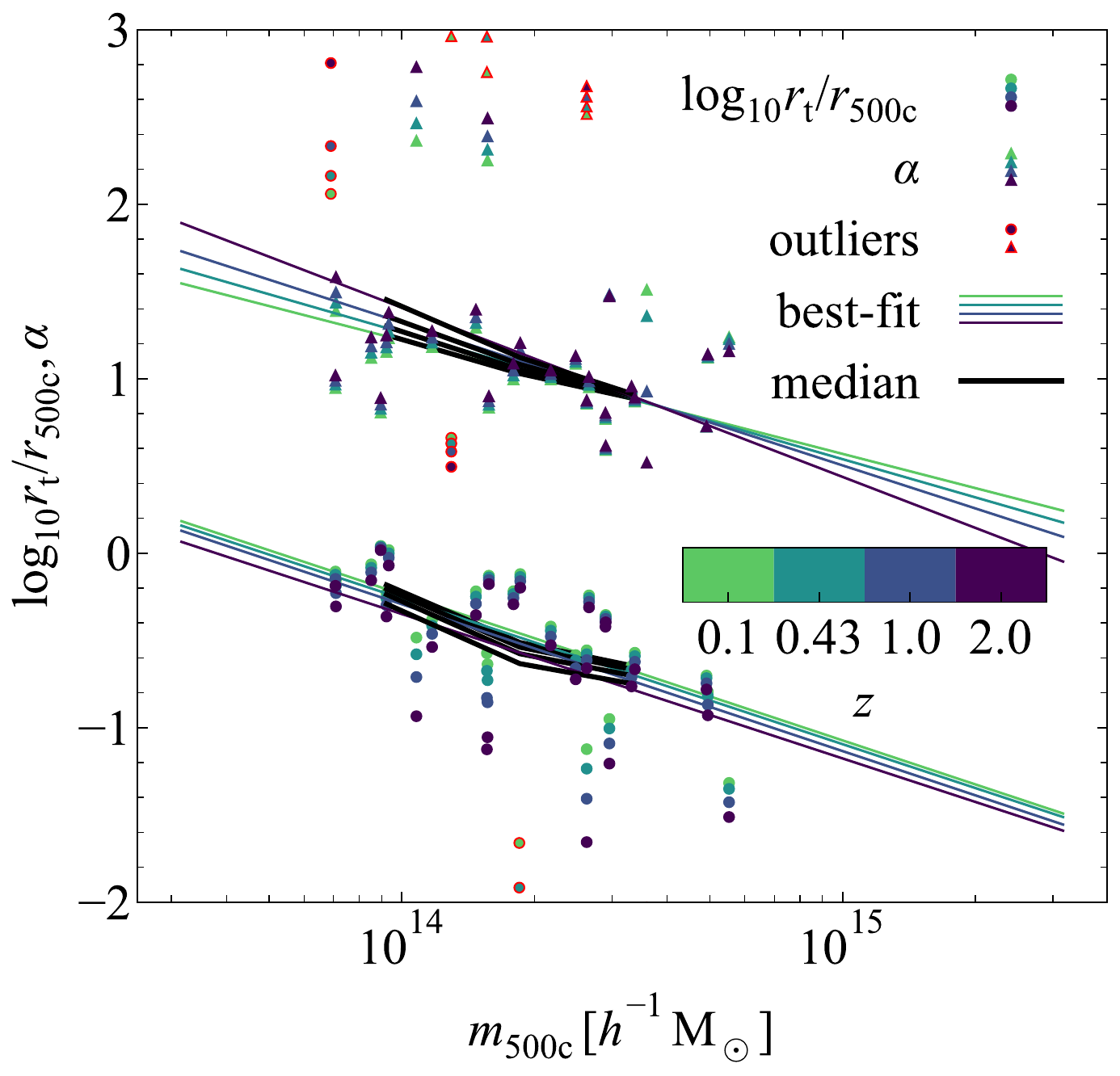}
  \caption{The best-fitting $\log_{10}(r_\mathrm{t}/r_\mathrm{500c})$ and
    $\alpha$ values for the REXCESS clusters, self-similarly rescaled to
    different redshifts (coloured points), their median, mass-binned values
    (black lines) and the best-fitting linear relations from
    Eqs.~\eqref{eq:model_rt} \& \eqref{eq:model_alpha} (coloured lines). The
    median relation is captured well with the linear model for each redshift.
    There are some outliers (red outlined markers), whose density profiles are
    shown in
    Fig.~\ref{fig:model_parameters_outliers}.}\label{fig:model_parameters}
\end{figure}
\begin{figure}
  \centering \includegraphics[width=\columnwidth]{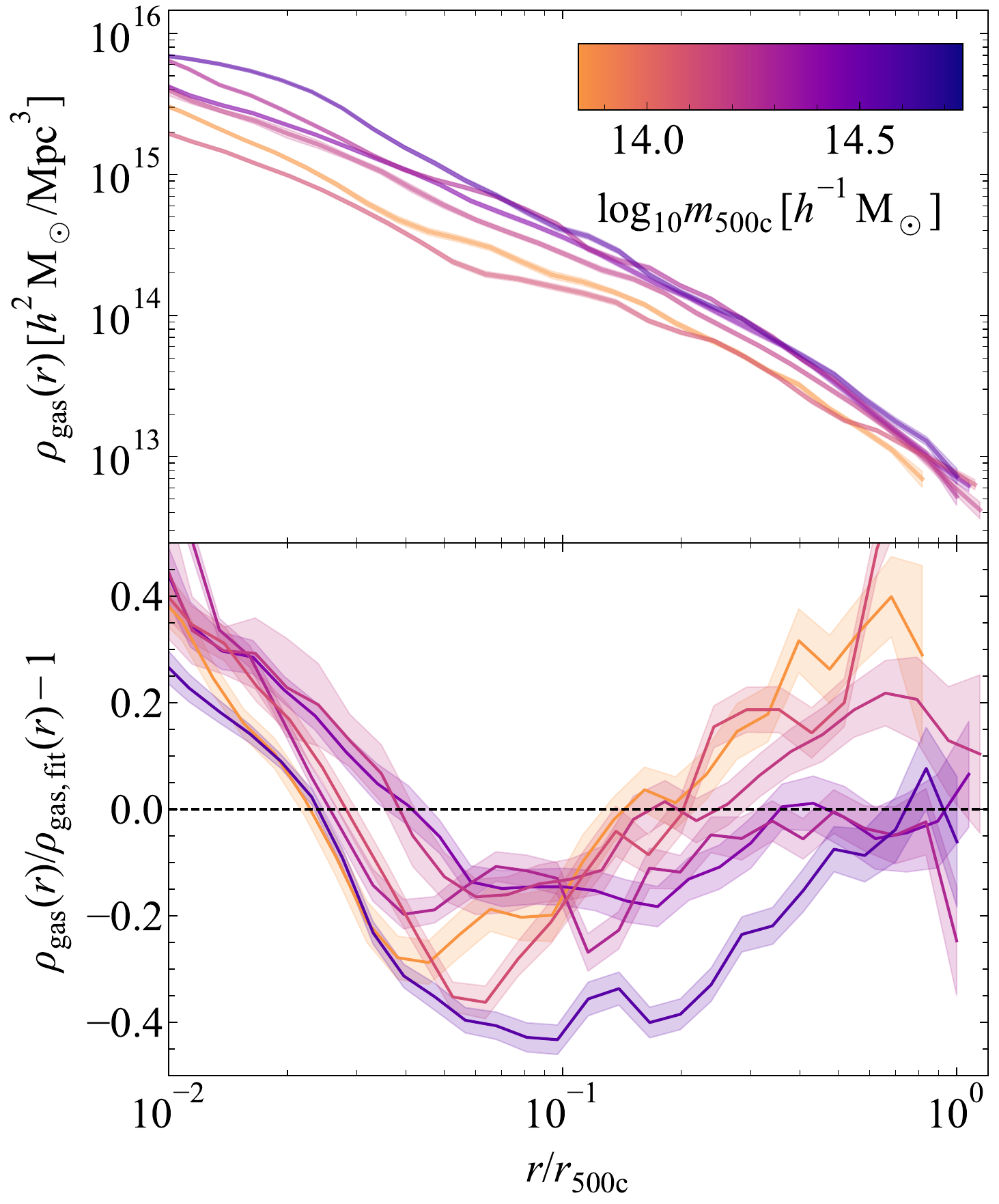}
  \caption{\emph{Top panel:} The density profiles of the clusters that are
    outliers to our best-fitting relations for
    $\log_{10}(r_\mathrm{t}/r_\mathrm{500c})$ and $\alpha$. \emph{Bottom panel:
    } The ratio between the observed hot gas density profiles and our
    best-fitting model. The outliers cannot be accurately described by our
    simple monotonic increase in the baryon fraction because they have a high
    density core.}\label{fig:model_parameters_outliers}
\end{figure}

Fig.~\ref{fig:model_parameters} shows the best-fitting
$r_\mathrm{t}(m_\mathrm{500c}, z)$ and $\alpha(m_\mathrm{500c}, z)$ for the
radial baryon fraction fits (Eq.~\ref{eq:fbar_fun}) to each cluster in the
REXCESS data, self-similarly scaled to the indicated redshifts. We also show the
results for the binned clusters as the black lines, and the best-fitting linear
relations, following Eqs.~\eqref{eq:model_rt} \& \eqref{eq:model_alpha}, as the
coloured lines. Most of the clusters are described quite well by the
best-fitting relations. In Fig.~\ref{fig:model_parameters_outliers}, we show the
outliers (marked in red in Fig.~\ref{fig:model_parameters}) with
$|\Delta \log_\mathrm{10}(r_\mathrm{t} / r_\mathrm{500c}) /
\log_\mathrm{10}(r_\mathrm{t} / r_\mathrm{500c})| > 1.5$ and
$|\Delta \alpha / \alpha| > 1.5$. All these clusters have a high central density
core that cannot be captured by our monotonic relation for the baryon fraction
(Eq.~\ref{eq:fbar_fun}). These clusters would be better described by, for
example, a double beta profile fit. However, these are only 6 out of the total
of 31 clusters, spanning the entire mass range. Hence, they do not bias the
median mass-binned cluster profiles.

\section{Mixed likelihood}\label{app:mixed_likelihood}

\begin{figure}
  \centering \includegraphics[width=\columnwidth]{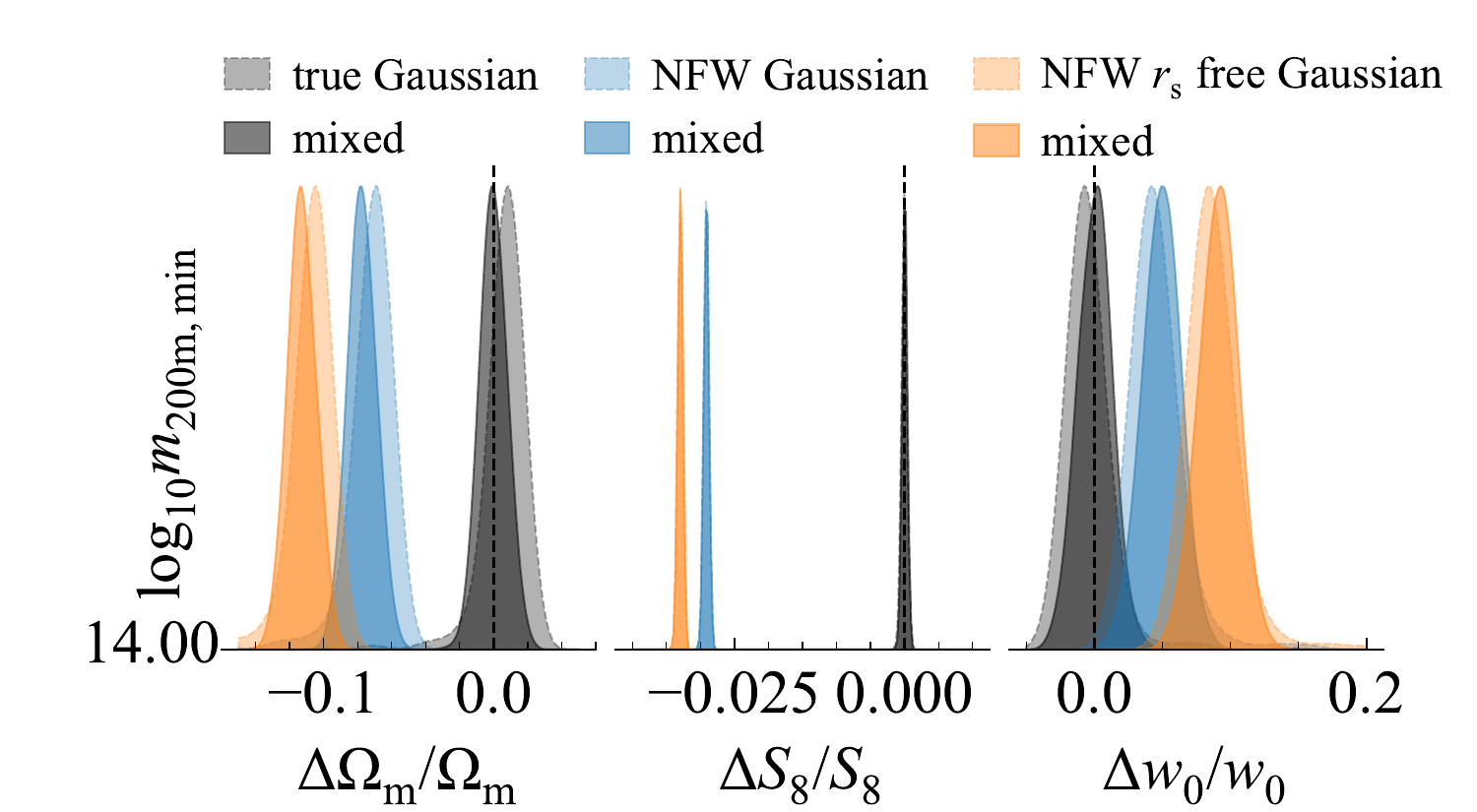}
  \caption{The marginalized maximum a-posteriori probability density functions
    for $\Omega_\mathrm{m}$, $S_8=\sigma_8(\Omega_\mathrm{m}/0.3)^{0.2}$ and
    $w_0$ for 1000 independent stage IV-like cluster abundance surveys assuming a
    Gaussian likelihood (lightly shaded contours), or a mixed Gaussian-Poisson
    likelihood (darkly shaded contours). Gray PDFs show the results for a halo
    sample with no mass bias. Blue (orange) PDFs include a mass bias due to an
    NFW fit with a fixed (free) scale radius, $r_\mathrm{s}$. The Gaussian
    likelihood biases $\Omega_\mathrm{m}$ ($w_0$) towards higher (lower)
    values.}\label{fig:cosmo_1d_likelihood}
\end{figure}

In Fig.~\ref{fig:cosmo_1d_likelihood} we show the difference in cosmological
parameter constraints for a stage IV-like cluster abundance survey when using a
pure Gaussian likelihood, i.e. Eq.~\eqref{eq:ln_like_gauss}, versus the mixed
Gaussian-Poisson likelihood that uses Eq.~\eqref{eq:ln_like_poisson_binned} for
bins with $N_\mathrm{obs}(m_i, z_j) < 10$. The Gaussian likelihood cannot deal
with the discreteness of the number counts at high redshift and high halo masses.
The absence of clusters in these bins pushes the theoretical prediction of the
halo mass function towards lower values in the Gaussian likelihood. Meanwhile,
the number counts for low-mass haloes, which are more abundant and thus better
described by the Gaussian likelihood, need to remain the same. For the mass cut
$m_\mathrm{200m,min} = 10^{14} \, \mh$, the lower number counts for high-mass
haloes are achieved by decreasing $w_0$ and increasing $\Omega_\mathrm{m}$ in
such a way that the decrease in number counts for low-mass haloes due to $w_0$ is
offset by the increase due to $\Omega_\mathrm{m}$. $S_8$ seems unaffected by the
choice in likelihood. The mixed Gaussian-Poisson likelihood results in unbiased
cosmological parameter estimates.

%%%%%%%%%%%%%%%%%%%%%%%%%%%%%%%%%%%%%%%%%%%%%%%%%%
% Don't change these lines
\bsp	% typesetting comment
\label{lastpage}
\end{document}